\PassOptionsToPackage{prologue, dvipsnames}{xcolor}
\documentclass[acmsmall,screen]{acmart}
\usepackage{amsmath,amsfonts}
\usepackage{algorithmic}
\usepackage{graphicx}
\usepackage{textcomp}
\usepackage[dvipsnames]{xcolor}
\def\BibTeX{{\rm B\kern-.05em{\sc i\kern-.025em b}\kern-.08em
    T\kern-.1667em\lower.7ex\hbox{E}\kern-.125emX}}
\usepackage{listings}
\usepackage[utf8]{inputenc}
\usepackage{color}
\usepackage{tabularray}
\usepackage{graphicx}
\usepackage{colortbl}
\usepackage{threeparttable}
\usepackage{multirow}
\usepackage{fancybox}
\usepackage{setspace}
\usepackage{caption}
\usepackage{subcaption}
\usepackage{algorithm}
\usepackage{algorithmic}
\usepackage[most]{tcolorbox}
\usepackage{adjustbox}
\usepackage{nicematrix}
\usepackage{multirow}
\usepackage{threeparttable}
\usepackage{hyperref}
\usepackage{balance}
\usepackage{enumitem}
\usepackage{tabularray}
\usepackage{booktabs}
\usepackage{xspace}
\usepackage{collab}

\def\name{\textsc{LogUpdater}\xspace}
\definecolor{codegreen}{rgb}{0,0.6,0}
\definecolor{codegray}{rgb}{0.5,0.5,0.5}
\definecolor{codepurple}{rgb}{0.58,0,0.82}
\definecolor{backcolour}{rgb}{0.95,0.95,0.92}
\definecolor{cerulean}{rgb}{0.0, 0.48, 0.65}
\definecolor{ceruleanblue}{rgb}{0.16, 0.32, 0.75}
\definecolor{cadmiumred}{rgb}{0.89, 0.0, 0.13}
\definecolor{grey}{rgb}{0.9, 0.9, 0.9}
\definecolor{viol}{RGB}{134,0,175}
\definecolor{githubgreen}{RGB}{204, 255, 204}
\definecolor{githubred}{RGB}{255, 224, 224}
\definecolor{mygray}{rgb}{0.8,0.8,0.8}
\definecolor{lightyellow}{rgb}{1,1,0.8}
\lstdefinestyle{test}{
    language={sh},
    moredelim=**[is][\color{red}]{~}{~},
    basicstyle=\ttfamily, 
}

\collabAuthor{ry}{red}{Renyi Zhong}
\collabAuthor{jx}{green}{Jinxi Kuang}
\collabAuthor{yc}{blue}{Yichen Li}
\collabAuthor{yt}{orange}{Yintong Huo}
\collabAuthor{ww}{violet}{Wenwei Gu}

\AtBeginDocument{%
  \providecommand\BibTeX{{%
    Bib\TeX}}}

\setcopyright{cc}
\setcctype{by}
\acmJournal{TOSEM}
\acmYear{2025} \acmVolume{1} \acmNumber{1} \acmArticle{1} \acmMonth{1} \acmPrice{}\acmDOI{10.1145/3731754}

\begin{document}

\title{\name: Automated Detection and Repair of Specific Defects in Logging Statements}

\author{Renyi Zhong}
\email{renyizhong@link.cuhk.edu.hk}
\affiliation{%
  \institution{The Chinese University of Hong Kong}
  \country{Hong Kong}
}

\author{Yichen Li}
\email{ycli21@cse.cuhk.edu.hk}
\affiliation{%
  \institution{The Chinese University of Hong Kong}
  \country{Hong Kong}
}

\author{Jinxi Kuang}
\email{jxkuang22@cse.cuhk.edu.hk}
\affiliation{%
  \institution{The Chinese University of Hong Kong}
  \country{Hong Kong}
}

\author{Wenwei Gu}
\email{wwgu21@cse.cuhk.edu.hk}
\affiliation{%
  \institution{The Chinese University of Hong Kong}
  \country{Hong Kong}
}

\author{Yintong Huo}
\authornote{Yintong Huo is the corresponding author.}
\email{ythuo@smu.edu.sg}
\affiliation{%
  \institution{Singapore Management University}
  \country{Singapore}
}

\author{Michael R. Lyu}
\email{lyu@cse.cuhk.edu.hk}
\affiliation{%
  \institution{The Chinese University of Hong Kong}
  \country{Hong Kong}
}


\begin{abstract}

Developers write logging statements to monitor software runtime behaviors and system state. However, poorly constructed or misleading log messages can inadvertently obfuscate actual program execution patterns, thereby impeding effective software maintenance.
Existing research on analyzing issues within logging statements is limited, primarily focusing on detecting a singular type of defect and relying on manual intervention for fixes rather than automated solutions.

To address the limitation, we initiate a systematic study that pinpoints four specific types of defects in logging statements (i.e., statement code inconsistency, static dynamic inconsistency, temporal relation inconsistency, and readability issues) through the analysis of real-world log-centric changes. We then propose \name, a two-stage framework for automatically detecting and updating logging statements for these specific defects.
In the offline stage, \name constructs a similarity-based classifier on a set of synthetic defective logging statements to identify specific defect types. 
During the online testing phase, this classifier first evaluates logging statements in a given code snippet to determine the necessity and type of improvements required.
Then, \name constructs type-aware prompts from historical logging update changes for an LLM-based recommendation framework to suggest updates addressing these specific defects.

We evaluate the effectiveness of \name on a dataset containing real-world logging changes, a synthetic dataset, and a new real-world project dataset. The results indicate that our approach is highly effective in detecting logging defects, achieving an F1 score of 0.625. 
Additionally, it exhibits significant improvements in suggesting precise static text and dynamic variables, with enhancements of 48.12\% and 24.90\%, respectively. 
Furthermore, \name achieves a 61.49\% success rate in recommending correct updates on new real-world projects. We reported 40 problematic logging statements and their fixes to GitHub via pull requests, resulting in 25 changes confirmed and merged across 11 different projects.

\end{abstract}


\maketitle

\section{Introduction}

Logging is a fundamental practice to provide visibility of systems status, especially for revealing complicated software activities in the production environments~\cite{he2021survey,jiang2024lilac}.
Listing~\ref{lst:example} shows a logging statement in if-block containing three ingredients: logging level (i.e., \textit{debug}), static contents (i.e., \textit{Left sub-range fully inconsistent}), and dynamic variables (i.e., \textit{depth} and \textit{right})~\cite{jiang2024large, huang2024ulog}.

When executed, these logging statements generate log messages that serve as primary resources for observing the behavior of systems during runtime. Such messages are instrumental in tasks like identifying anomalies~\cite{liu2023scalable,yu2024deep,zhang2024metalog} and probing into root cause analyses~\cite{li2024exchain,chen2024automatic}. Therefore, the accuracy and clarity of these logging statements are crucial for providing reliable system information to engineers. However, logging statements containing defects—such as inconsistencies in semantics, incorrect variable usage, or tense mismatches—can hinder their monitoring purpose and result in the dissemination of erroneous information. This, in turn, can prolong error diagnosis procedures and delay anomaly resolution~\cite{yang2024try}. While logging defects encompasses various aspects, including appropriate log levels and information verbosity, the defect mentioned in the following paper focuses specifically on factual defects related to the content and context of logging statements.
Taking the above Cassandra code in Listing 1 as an example\footnote{\url{https://issues.apache.org/jira/browse/CASSANDRA-15173}}, the static content of this logging statement exhibits a semantic inconsistency with the dynamic variable: the static content indicates the left tree is recorded, but the dynamic variable actually represents the right tree. The corresponding produced log message will mislead engineers into looking at the wrong tree when executed.

\begin{lstlisting}[language=Java,showstringspaces=false,caption={A logging statement example from Cassandra.},label={lst:example},basicstyle=\small\ttfamily]
static int differenceHelper(...) { ...
  else if (!lreso) {
    log.debug("({}) Left sub-range fully inconsistent {}", depth, right);
    ldiff = FULLY_INCONSISTENT;
\end{lstlisting}

Despite the prevalence of logging in modern software systems, the issues in existing logging statements receive insufficient attention due to the following challenges~\cite{gholamian2021comprehensive}. Firstly, the large volume of logging statements within a codebase makes manual inspection impractical~\cite{10.1145/2110356.2110360,chen2024tracemesh}. 
Secondly, because defects in logging statements do not directly affect program functionality, the functional testing suites fail to detect such issues~\cite{bouzenia2023When}. 
Consequently, logging statements written by individual developers may not undergo a thorough review during the development and deployment process~\cite{zhuLearningLogHelping2015}.

While most studies have developed tools to decide logging locations~\cite{zhaoLog20FullyAutomated2017,liWhereShallWe2020,candidoExploratoryStudyLog2021} and suggest logging content~\cite{dingLoGenTextAutomaticallyGenerating2022,liuWhichVariablesShould2019,mizouchiPADLADynamicLog2019, li2024go, xu2024unilog,liDeepLVSuggestingLog2021}, only a few have looked into identifying defective logging statements in existing software systems~\cite{ding2023Temporal,bouzenia2023When,chen2017Characterizinga}. For instance, Ding et al.~\cite{ding2023Temporal} studied the temporal relation between logging statements and the actual code sequence, proposing a detection framework for log-code temporal issues. Li et al.~\cite{li2023Are} analyzed the readability of logging statements and attempted an automated readability issue detection process to mitigate it.
However, existing works suffer from two main drawbacks: \textbf{(1) Limited scope of analysis:} 
Previous studies focused on a single, specific category of logging statement defect (e.g., exclusively detect temporal inconsistencies~\cite{ding2023Temporal} or only focus on readability issues~\cite{li2023Are}), failing to provide a comprehensive view of logging statement defects. While combining the outputs of these specialized tools is theoretically possible, this approach presents practical limitations. Firstly, it requires developers to manage and execute multiple disparate tools, increasing workflow complexity and potentially analysis time. Secondly, integrating results from different tools for a unified diagnosis can be cumbersome. Most importantly, such a combination does not readily lend itself to a unified automated repair mechanism. Effective automated repair often requires understanding the specific type of defect present to apply the correct fixing strategy. Building a cohesive, type-aware repair engine on top of independent detectors is a significant challenge.
\textbf{(2) Detection without repairing:}
Existing works solely focused on detecting certain defects in logging statements without offering suggestions for repairs. Therefore, these approaches require extra human intervention with domain-specific knowledge to correct the defects.
Automating the process of updating logging statements remains an unexplored area.

To tackle the first limitation, we begin with a pilot study to uncover the potential defects that can affect the accuracy and clarity of logging statements by analyzing the existing logging-related software updates. Unlike previous studies~\cite{ding2023Temporal,chen2017Characterizinga} focused on a small number of projects (\textasciitilde10), our pilot study spans 641 repositories, providing a broader range to analyze defective logging practices across various real-world software development environments. Upon analysis, we identify four distinct objective defects, each requiring precise detection and \textit{defect-specific remediation strategies}.
In addition to our pilot study findings, through a survey of related work, we observe two important considerations that inform our approach:
(1) while defective logging statements frequently appear in logging-related updates, they represent only a minor fraction of the overall logging statement corpus. This observation is supported by multiple empirical studies: Kabinna et al.\cite{kabinna2016Examining} found that only 30\% of logging statements undergo changes during their lifetime, while Chen et al.\cite{chen2017characterizingb} reported that approximately 60\% of logging statement changes are after-thought modifications not directly related to core feature implementation, with only about a third addressing misleading content. This \textit{imbalanced distribution of data} highlights the need to \textit{consider the resource overhead} of our approach, particularly when employing large language models (LLM) for fixing; and
(2) recent research shows that learning-based methods can detect defects in certain categories of logging statements~\cite{bouzenia2023When,ding2023Temporal}. Additionally, repairing defects with contextual defect information yields better results compared to arbitrary fixing~\cite{yang2023significance}. Therefore, an effective strategy involves designing a \textit{defect detector, followed by suggesting repair updates} based on the detection results.

Based on the above insights and to address the second limitation, we propose \name, a framework for automatically detecting and updating defective logging statements. \name contains an offline fine-tuning phase and an online phase for updating based on detection.  
The offline part aims to develop a defective logging statement detector. For the four distinct defect types, we first design three efficient mutation strategies and then fine-tune a similarity-based classifier for defect type identification on a set of synthetic defective logging statements. 
During the online phase, \name first applies the above classifier to determine whether the input code snippet needs improvement and identify the specific types of defects.
Finally, \name constructs defect type-aware prompts and incorporates surrounding code contexts for suggesting logging statement updates.

We extensively evaluate \name in three datasets, including an existing log patches dataset collected from real-world software history, a synthetic dataset containing artificial defects, and a new dataset with unidentified defects for human evaluation.
Our approach achieves promising detection ability (0.625 at F1) and a significant boost of 48.12\% and 24.90\% in static text and dynamic variables update ability, respectively. \name also demonstrates effective detection and updating capabilities on new project data, achieving an overall successful rate of 61.49\%. To date, 25 out of 40 changes submitted to the GitHub project developers have been merged, underlining the practicality of \name.

In summary, this paper makes the following contributions:
\begin{itemize}[leftmargin=*]
    \item We conduct a broad-scope pilot study spanning 641 repositories, surpassing previous studies that focused on specific libraries. Our broad-scope analysis of defective logging practices across diverse development environments identify four types of common defects that developers are concerned about, providing more representative defects for real-world software.
    \item We propose \name, a novel end-to-end framework that automatically detects and repairs four categories of defective logging statements. Unlike existing tools limited to specific defect types, our solution addresses a wider range of defects, effectively identifying subtle and complex issues often missed by category-specific approaches.
    \item Experimental validation across three datasets, including successful merging of 25 PRs in popular projects demonstrates our framework's practical effectiveness. The tool's automated repair capability overcomes manual maintenance limitations in large-scale projects. To facilitate reproducibility, we open-sourced relevant code, prompts, and datasets~\cite{package}.
\end{itemize}

\textbf{Paper Organization}
Section~\ref{motivation} conducts a pilot study to investigate the defects in logging statements. Section~\ref{approach} introduces our framework for automating detecting and updating defective logging statements. Section~\ref{implementation} shows the implementation details. Section~\ref{evaluation} describes the experimental results. Section~\ref{discussion} presents a discussion on the practicality of \name and threats to validity. Section~\ref{related_work} discusses the related work. Section~\ref{conclusion} concludes the paper.

\section{Pilot Study}
\label{motivation}
In this section, we investigate the defects in existing logging statements in real-world software applications. Considering that modifications to logging statements typically occur concurrently with functional code updates~\cite{yuanCharacterizingLoggingPractices2012}, we posit that isolated commit changes related to existing statements are likely aimed at addressing defect issues. Consequently, our analysis centres around examining what is referred to as the \textit{log-centric change (i.e., LCC)}. This term describes software commits that exclusively involve the modification, excluding both addition and deletion, of one or more logging statements without accompanying modifications to functional code. Similar to the approach of related studies~\cite{chen2017Characterizinga}, we exclude the addition and deletion of logging statements from our analysis. This exclusion is primarily due to the unique challenges involved in modifying existing logging statements, which necessitate an analysis of semantic alignment between code and logs. In contrast, scenarios involving the addition of logging statements require distinct analytical frameworks to pinpoint potential logging locations~\cite{liWhereShallWe2020}, while situations involving the deletion of these statements raise separate concerns, such as redundancy~\cite{liDLFinderCharacterizingDetecting2019}.

\subsection{Dataset Construction}
The majority of current research works~\cite{ding2023Temporal,chen2019extracting,liDLFinderCharacterizingDetecting2019,li2021studying} predominantly focus their investigations on a handful, specifically between four to six, of well-established repositories such as Hadoop and Zookeeper. This narrow focus results in a lack of an inclusive perspective on logging statements within newer and developing software systems.
In contrast, our preliminary study is intended to encompass a larger number of repositories, with a broader domain, lifecycle, and scale, thereby providing a more diverse study.

\subsubsection{Select target repositories}
We first collect projects from popular and developing Java repositories on Github, denoted as target repositories.
Specifically, the target repositories satisfy the following criteria~\cite{wen2019large}:
\begin{itemize}[leftmargin=*]
    \item The number of commits for the projects should range between 1,000 and 100,000, ensuring sufficient community interest and maintenance.
    \item Projects must have been created before December 31, 2019, and the last commit must be later than January 1, 2023. This criterion ensures that the projects have a significant history and sustained maintenance.
    \item The number of stars for the projects must exceed 1,000, indicating the popularity of the projects.
\end{itemize}

We utilized the tool provided by \citet{Dabic:msr2021data} to filter GitHub repositories based on the aforementioned criteria. After excluding forks and merging repositories with the same name from different contributors, we obtained a total of 749 repositories that met the specified conditions. To ensure that each target repository contains at least one logging statement, we employed a static analysis method based on JavaParser to decompose the corresponding Java class files into units of methods. This process allowed us to eliminate repositories that did not contain any logging statements. Finally, we obtain a total of 641 repositories as the target datasets. Table~\ref{tab:dataset} shows statistics of these repositories.

\begin{table}[t]
\centering
\small
\caption{Statistics of the studied repositories.}
\label{tab:dataset}
\begin{tabular}{c|cccc}
\toprule
Avg \# & Code line & Fork  & Commit & Issue \\ \toprule
Target repositories       & 454,157   & 1,886 & 7,618  & 1,772 \\ \bottomrule
\end{tabular}
\end{table}

\subsubsection{Extracting log-centric changes}
We then track changes in logging statements based on the evolution history of the target repository.
In specific, we start with developing scripts to execute the git log command on the historical codebase from target repositories. These scripts extract the source code and relevant metadata for each commit, including identifiers such as the commit hash and content. 
Then, we employ ChangeDistiller~\cite{fluri2007change} to localize modifications between consecutive versions. 
Furthermore, considering that logging statements typically describe the code context within a method~\cite{fuWhereDevelopersLog2014}, we isolate functions containing logging statements from the repository. This aligns with previous intra-function logging practices research settings~\cite{yuanCharacterizingLoggingPractices2012,shang2015studying}. 
Next, we filter out the commits including functional code changes, reserving those where all modifications are related to logging lines. Ultimately, we identified 10,137 LCCs.

\subsection{Manual Study Design}
\label{manualstudy}

To execute a comprehensive manual analysis, we adhere to the sampling methodology outlined in prior research efforts~\cite{chen2018automated,li2023exploring}. In detail, we apply a random sampling technique~\cite{pirzadeh2011concept} for the selection of 500 instances of LCC, ensuring a 95\% confidence level and a margin of error at 5\% confidence interval~\cite{boslaugh2012statistics}. We enlist two proficient co-authors, referred to as annotators A1 and A2, who each possess more than seven years of programming expertise. Given the intricate nature of this research, we administer a two-stage examination procedure. This dual-phase approach enables us to ultimately obtain 500 LCCs, each manually categorized according to their respective defect types.

Stage 1: Identifying categories of defect. In this stage, we first randomly select 100 logging statements with their surrounding code, where A1 and A2 first derive a draft list of categories of logging statements separately. Then they discuss the different cases until reaching a consensus. During this phase, the categories are revised and refined.

Stage 2: Categorizing all sampled data. In this stage, A1 and A2 independently categorize the remaining 400 logging statements by investigating the reason for changes by analyzing the data flow and structural context surrounding each logging statement. After this, their categorizations are compared. When both agree on a label, it's accepted as the final category. Before any discussion, the results yield a Cohen's Kappa of 0.78, which is a substantial level of agreement. However, when there are disagreements, the two annotators discuss collaboratively to decide on the final label for the disputed logging statements. Ultimately, they come to a mutual agreement on all categorizations.

\begin{table}[t]
\centering
\small
\caption{Summarized rationals for log-centric changes.}
\label{tab:manual_res}
\begin{tabular}{l|cc}
\toprule
\multicolumn{1}{c}{Type}              & \multicolumn{1}{c}{Number} & \multicolumn{1}{c}{Percentage} \\ \toprule
Correct statement-code inconsis.    & 73  & 14.6\% \\
Correct static-dynamic inconsis.  & 45  & 9.0\%  \\
Correct temporal inconsis.        & 11  & 2.2\%  \\
Correct typos/capitalization & 129                        & 25.8\%                         \\
 \midrule
Add/delete statement info & 105 & 21.0\% \\
Change log level          & 124 & 24.8\% \\
Refactor output format    & 8   & 1.6\%  \\
Others                    & 5   & 1.0\%  \\ \midrule
Total                     & 500 & 100\% 
\\ \bottomrule
\end{tabular}
\end{table}

\subsection{Defects Type in Log-centric Changes}

Table~\ref{tab:manual_res} provides an overview of the results obtained from the manual analysis. This table outlines seven distinct reasons for updating logging statements.
Our research specifically concentrates on the first four categories of these reasons because they relate to factual defects. These defects result in errors in the objective description of system behaviors and have a direct influence on the accuracy and clarity of the logging statements. We have intentionally omitted subjective alterations and project-specific conventions from our focus. By doing so, we aim to concentrate on defects that have verifiable consequences on logging statements, thereby enhancing the generalizability of our taxonomy. This approach helps us to avoid merging stylistic preferences with factual inaccuracies. 
The choice to \textit{add or delete statement information} relies heavily on subjective interpretations concerning what is deemed as "sufficient" information. As our methodology emphasizes objective and verifiable defects, the inclusion of highly subjective categories could jeopardize the uniformity and dependability of the processes for detecting and correcting these issues. 
Alterations in \textit{log levels}, such as changing a log level from debug to info, can exhibit considerable variation between projects. These changes are often driven by organizational or project-specific logging methods. This degree of variability poses a significant challenge in defining a universal criterion for identifying what constitutes a defect in log-level settings, which is why it remains outside the scope of our main focus.
Notably, we observe that 51.6\% (258/500) of the changes relate to the first four factual defects, highlighting the significant attention developers devote to factual defect issues.
More specifically, we detail the symptoms and explain how each defect weakens the accuracy and clarity of logging statements as follows.  Figure~\ref{fig:defects_example} offers examples of each defect from four different repositories.

\subsubsection{Statement-code inconsistency}
This defect indicates that a logging statement whose semantic meaning conflicts with the execution logic within its enclosing method.
The fundamental purpose of logging practice is to capture the runtime system activities of the current program for operation.
A mismatch between the content of the logging statements and their code context leads to inconsistencies that can mislead engineers in failure diagnosis. For example, in Fig.~\ref{fig:example1}, the \textit{DDL success} event was mistakenly logged as \textit{Catalog update success}.

\subsubsection{Static-dynamic inconsistency}
\begin{figure}[tbp]
    \centering
    \begin{subfigure}[b]{0.85\columnwidth}
        \centering
        \includegraphics[width=\columnwidth]{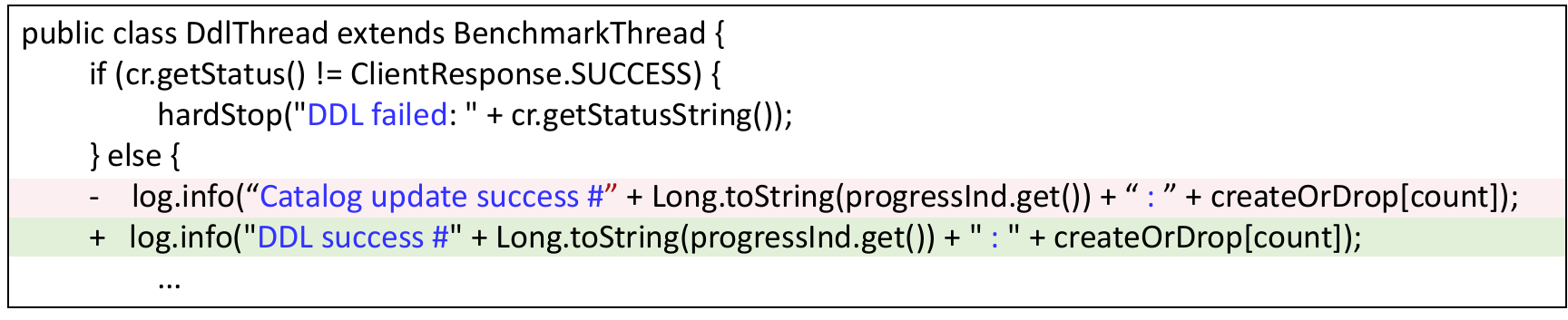}
        \caption{Statement code inconsistency at \href{https://github.com/VoltDB/voltdb/commit/f5da74bcb0ae03a616ff1c75a986cd23a1ea44ac}{DdlThread.java} in voltdb.}
        \label{fig:example1}
    \end{subfigure}
    \hfill
    \begin{subfigure}[b]{0.85\columnwidth}
        \centering
        \includegraphics[width=\columnwidth]{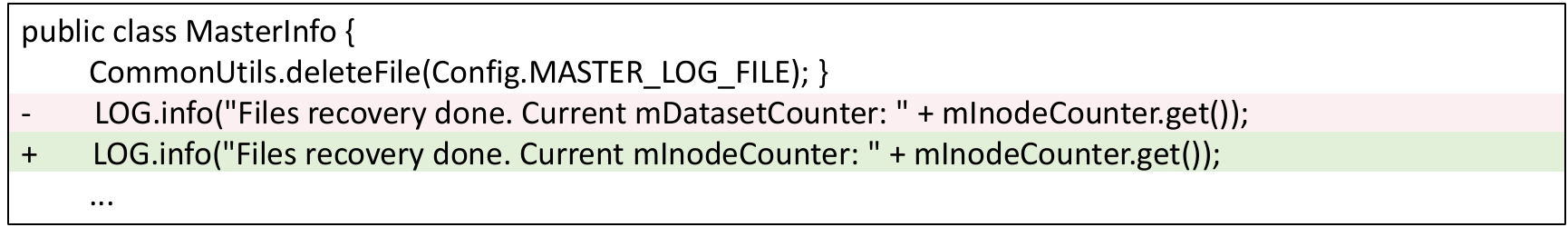}
        \caption{Static dynamic inconsistency at \href{https://reurl.cc/E67VgK}{MasterInfo.java} in alluxio.}
        \label{fig:example2}
    \end{subfigure}
    \vfill
    \begin{subfigure}[b]{0.85\columnwidth}
        \centering
        \includegraphics[width=\columnwidth]{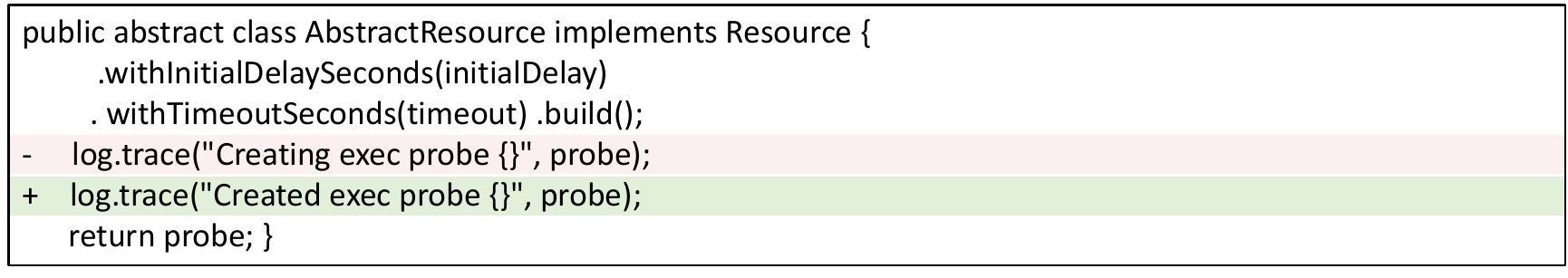}
        \caption{Temporal relation inconsistency at \href{https://github.com/strimzi/strimzi-kafka-operator/commit/62067caf5a77c8982560b688501b5e2d7bae49ba}{AbstractResource.java} in strimzi-kafka-operator.}
        \label{fig:example3}
    \end{subfigure}
    \hfill
    \begin{subfigure}[b]{0.85\columnwidth}
        \centering
        \includegraphics[width=\columnwidth]{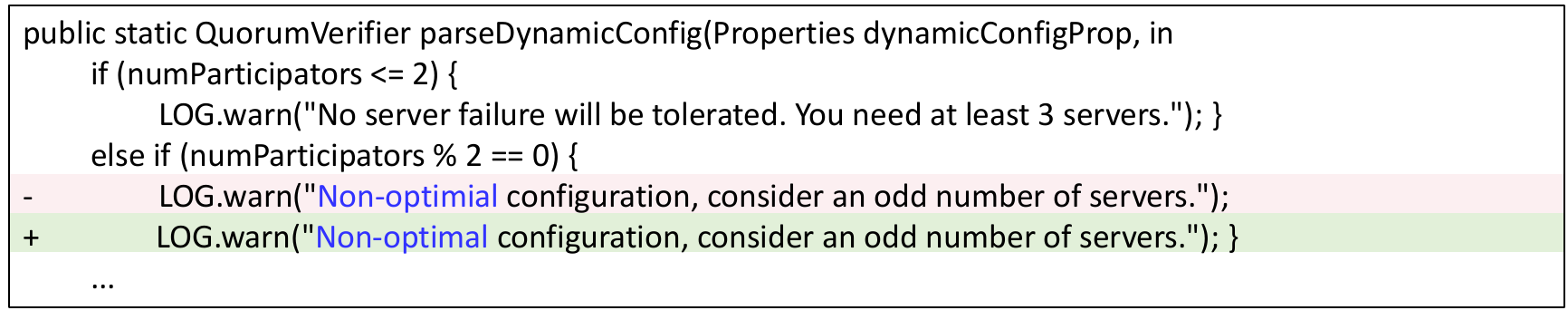}
        \caption{Readability issue at \href{https://github.com/apache/zookeeper/commit/3b6fefc43221fb3626740618a68562ff1ba707c0}{QuorumPeerConfig.java} in zookeeper.}
        \label{fig:example4}
    \end{subfigure}
    \caption{Real-world examples for four defects.}
    \label{fig:defects_example}
\end{figure}

It represents a mismatch between the static text description and the actual content of dynamic variables. The static-dynamic inconsistency is considered if either: variable's name/usage in code contradicts the static text description (e.g., logging "Counter A: {}" while passing counter B variable) or the variable is missing while static text requires one (e.g. logging three variables in text but give two variables).
Static text within logging statements must carry the meaning and implications of the associated dynamic variables. Otherwise, the logged variable strings are meaningless and hard to comprehend for log inspectors. As in Fig.~\ref{fig:example2}, the static text \textit{``Current mDatasetCounter: ''} incorrectly describes the dynamic variable \textit{mInodeCounter.get()}. This misalignment could lead to confusion when debugging or reviewing logs.

\subsubsection{Temporal inconsistency} 
To address temporal inconsistency, we adhere to the definition proposed by Ding et al.~\cite{ding2023Temporal}. This definition characterizes temporal inconsistency as arising when there is a mismatch between the logical temporal relations, which pertain to the sequence in which a logging statement is executed in relation to its associated source code (i.e., what the actual order is in the code), and the semantic temporal relations, which can be deduced from the semantic interpretation of the logging text (i.e., what the order is inferred from the generated logs).
The tense used in logging statements should align with the temporal context of the events. This adherence ensures clarity for tracking system events.
In Fig.~\ref{fig:example3}, the original logging statement uses the present continuous tense \textit{``Creating''}, which implies that the action of creating the \textit{exec probe} is still ongoing. However, the logging statement is placed immediately before the return statement, indicating that the creation of the probe has been completed by the time of the logging statement execution.

\subsubsection{Readability issues}
Readability issues are linguistic defects that impairing logging message clarity, which includes spelling errors and capitalization errors.
Eliminating these errors is crucial to prevent misunderstandings and ensure that logs serve as clear and effective communication tools for developers and operators. Taking Fig.~\ref{fig:example4} as an example, \textit{"Non-optimial"} is a typo error. The readability issue of logging statements is significant since many logs generated by logging statements are processed by automated tools for downstream tasks like anomaly detection and root cause analysis. While current models used in these tools can often tolerate minor typos, such defects can still degrade the performance of embeddings derived from logs, ultimately impacting the accuracy of automated analyses.

\begin{tcolorbox}[boxsep=1pt,left=2pt,right=2pt,top=3pt,bottom=2pt,width=\columnwidth,colback=white!95!black,boxrule=1pt, colbacktitle=white!30!black,toptitle=2pt,bottomtitle=1pt,opacitybacktitle=0.4]
\textbf{Summary:} 
The study uncovers four categories of factual defects (statement-code inconsistency, static-dynamic inconsistency, temporal inconsistency, and readability issues), which account for 51.6\% log-centric commits.
\end{tcolorbox}

\section{Framework}
\label{approach}

\subsection{Overview}

This paper resolves the defective logging statement detection and updating task, described as follows: 
Let $s_{\text{original}}$ denote the given target statement. The goal is to first classify $s_{\text{original}}$ into one of the following categories using a multi-class classification model $f$. This classification can be represented as:$f(s_{\text{original}}) = type_s$, where $type_s$ belongs to the set \{non-defect, $d_{\text{statement\_code}}$, $d_{\text{static\_dynamic}}$, $d_{\text{temporal}}$, $d_{\text{readability}}$\}. If $type_s \neq \text{non-defect}$, the target statement $s_{\text{original}}$ is updated to obtain an improved statement $s_{\text{update}}$.

\begin{figure*}[t]
    \centering
    \includegraphics[width=0.99\textwidth]{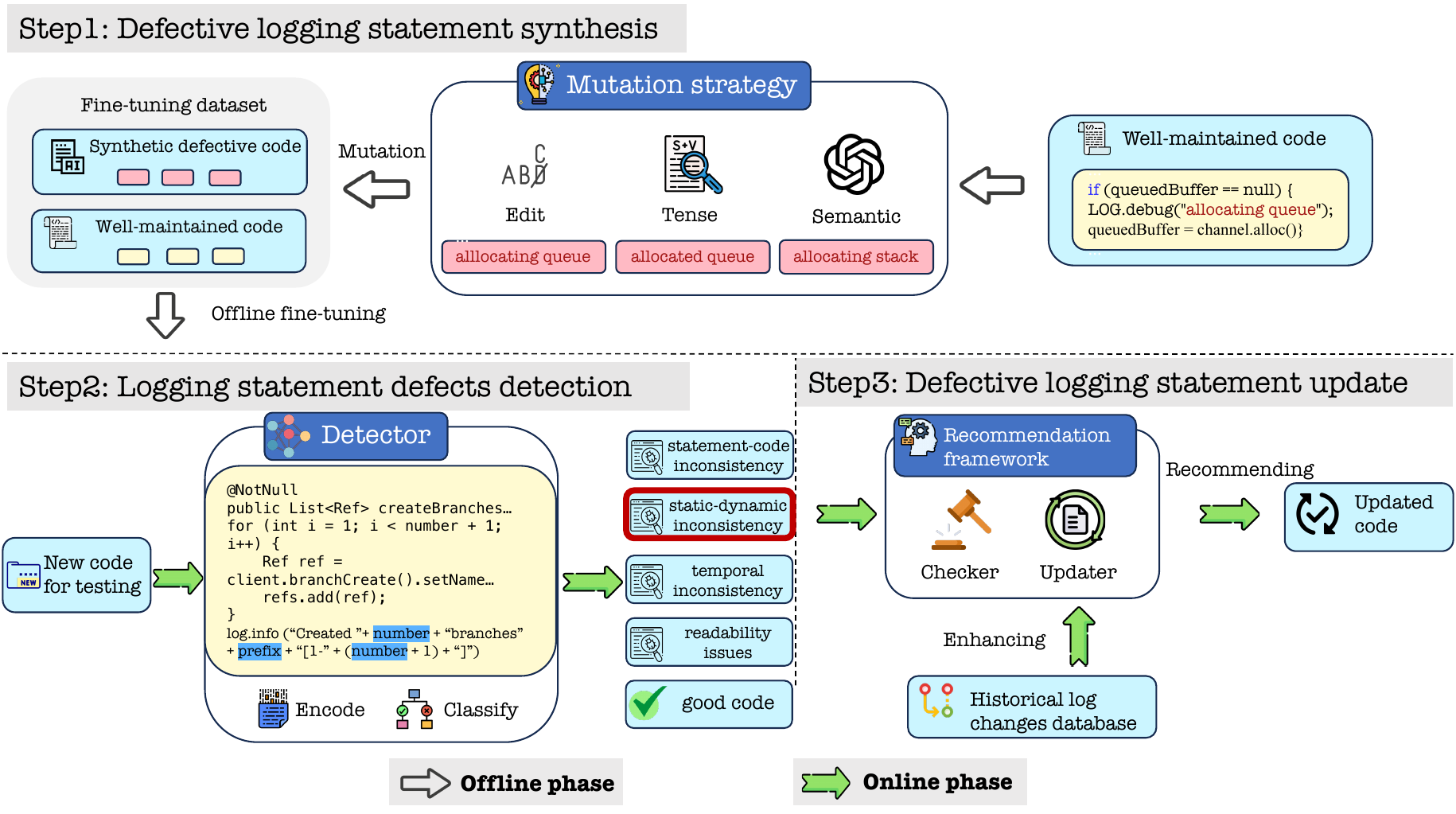}
    \caption{The overview framework of \name.}
    \label{fig:framework}
\end{figure*}

To this end, we propose \name, a framework for automatically detecting and updating defective logging statements. Fig.~\ref{fig:framework} illustrates the high-level view of \name, which operates in two stages: offline fine-tuning and online updating. 
During the \textbf{offline stage}, the goal is to develop a defective logging statement detector. We first synthesize the defective logging statements with corpus collected from well-maintained repositories, ensuring high-quality code and maintenance standards. 
Then we use the synthetic data to fine-tune a similarity-based logging defect detector. 
The \textbf{online phase} involves two steps: detecting defective logging statements and updating them.
We first use the fine-tuned detector to determine the $type_s$. Based on the defect information, we apply the analyzed knowledge into the LLM-based updater for logging statement update.

\subsection{Defective Logging Statement Synthesis}
\label{generate_data}

To prepare the corpus for fine-tuning the detector, the initial step involves synthesizing negative samples. These samples are sourced from well-maintained repositories and mutated by text mutation strategies, as the corpus for tuning defective logging statement detectors. 
This foundational step is crucial as the training data quality employed in developing the model significantly influences its overall performance. 
Previous studies~\cite{bouzenia2023When} emphasize three crucial objectives in the data generation methodology: \textit{realism}, \textit{diversity}, and \textit{scalability}. \textit{Realism} ensures that the dataset mirrors actual errors that developers might encounter. \textit{Diversity} is vital for enabling the model to recognize various types of issues. Lastly, \textit{scalability} is necessary since training a robust model typically demands a large dataset.
In alignment with these objectives, we develop three techniques to synthesize defective logging statements for different defects, detailed as follows.

\subsubsection{Mutation of words} 
For \textbf{readability issues}, we synthesize defective logging statements by making straightforward modifications to one word in the static constant part of the logging statement. Specifically, for simulating typos, we randomly select a word and alter its spelling based on the known typo dataset~\cite{hagiwara-mita-2020-github, MicrosoftMLServer2016}. If the chosen word is not in the dataset, we then randomly add, delete, or change one of its characters. For issues with capitalization abuse, we randomly select a word and convert it to uppercase. We ensure that only one type of word mutation (e.g., \textit{executor} $\rightarrow$ \textit{executer}) is applied to each logging statement in modification.

\begin{lstlisting}[language=Java,showstringspaces=false,basicstyle=\small\ttfamily,caption={A word mutation case for readability issue.}, captionpos=t]
public void stop() {
    for (InstanceExecutor executor : Executors){ 
    executor.stop();}
    LOG.info("To stop Stream executer");
    // *** executor -> executer ***
    streamExecutor.stop();
\end{lstlisting}

\subsubsection{Mutation of tense}
Inspired by previous work~\cite{ding2023Temporal}, for \textbf{temporal inconsistency}, we use spaCy~\cite{Honnibal_spaCy_Industrial-strength_Natural_2020}, which is an open-source NLP library, to alter the tense of logging statements. Specifically, we employ dependency parsing~\cite{zmigrod-etal-2021-finding} and part-of-speech (POS) tagging~\cite{coleman2005introducing} to identify the main verbs in the logging statements. We consider the verb tagged as a head token in the dependency tree as the main verb. Finally, we change the main verbs to random tenses to mutate examples where logging statements do not align with the system events. For instance, we mutate \textit{starting} to \textit{started} before 
the sequence generator source starts in the following examples.

\begin{lstlisting}[language=Java,showstringspaces=false,basicstyle=\small\ttfamily,caption={A tense mutation case for temporal inconsistency.}, captionpos=t]
protected void doStart() throws FlumeException {
    logger.info("Sequence generator source do started");
    // *** starting -> started ***
    sourceCounter.start(); }
\end{lstlisting}

\subsubsection{LLM-based mutation} 
For defects requiring semantic mutation (\textbf{statement-code inconsistency} and \textbf{static-dynamic inconsistency}), using rule-based or heuristic methods can be challenging. Therefore, we employ LLM-based techniques to synthesize defective logging statements. These LLMs, trained on vast amounts of code, have been shown to handle code-related downstream tasks effectively~\cite{zhang2024code,xia2023automated,yang2023significance}. We use GPT-4o~\cite{OpenAIGPT4o} to mutate the semantics for static content or dynamic variables. Specifically, for \textit{statement-code inconsistency}, we prompt the LLM to carefully analyze the surrounding code and then alter the static content of logging statements to create semantic inconsistencies (e.g., recording \textit{opened} while the actual event is \textit{close}). For \textit{static-dynamic inconsistency}, we instruct the LLM to randomly mutate the static content or dynamic variables to generate potential mismatches (e.g., recording \textit{exc} while a directory path \textit{dir} is expected). The prompt details are shown in~\cite{mutationprompt}.

\begin{lstlisting}[language=Java,showstringspaces=false,basicstyle=\small\ttfamily,caption={A LLM-based mutation case for statement-code inconsistency.}, captionpos=t]
if (channelRef.get() != null) {
  channelRef.get().close();
  LOG.debug("channel {} opened", remoteAddr); 
  // *** closed -> opened *** }
\end{lstlisting}

\begin{lstlisting}[language=Java,showstringspaces=false,basicstyle=\small\ttfamily,caption={A LLM-based mutation case for static-dynamic inconsistency.}, captionpos=t]
public postVisitDirectory(Path dir, Exception exc) {
  LOGGER.info("> Deleting folder {}", exc);
  // *** dir -> exc ***
  Files.delete(dir); }
\end{lstlisting}

After generating defective logging statements using above techniques, we filter out the identical samples. This step is crucial as various data mutation techniques may inadvertently generate identical data. 
The presence of such duplicate data could potentially skew our detector's learning process, leading to an overemphasis on homogenous data features.

\subsection{Logging Statement Defects Detection}
Considering that only a small portion of logging statements require update, and scanning all methods in a repository with LLM is resource-intensive, \name employs a similarity-based classifier as the detector. This detector identifies defective logging statements and determines the types of defects. 
The introduction of the fine-tuned detector significantly reduces costs and enables subsequent LLM components to update the logging statements in a more efficient and targeted manner.
Specifically, this phase is responsible for handling a multi-class classification task. Given a logging statement $s_{original}$ with its associated context (i.e., code snippet), the goal is to determine whether $s_{original}$ has a specific type of defect. We resolve the problem through the following steps:

\subsubsection{Tokenization and embedding}
This step converts the input code snippet into a series of tokens.
In this study, we use UniXcoder~\cite{guoUniXcoderUnifiedCrossModal2022}, which is a unified cross-modal pre-trained model for programming languages that are based on a multi-layer transformer and utilizes mask attention matrices~\cite{dong2019unified} with prefix adapters to control the access to context for each token.
We adapt the pre-trained tokenizer provided by the Unixcoder and embed the series of tokens into fixed-size vectors. The fine-tuning process is end-to-end where the parameters of the encoder layers are adjusted to enhance the representation vectors, allowing these vectors to better adapt to detection task.

\subsubsection{Similarity-based multiple classification}
After obtaining the corresponding representations for the target method and logging statement, we then devise a multi-class classifier to identify defects.  Intuitively, a ``good'' logging statement should closely align with the semantic meaning of its adjacent code.
Hence, we have developed a similarity-based classification model instead of applying a basic multi-classifier with fully connected layers.
Specifically, the classifier concatenates the embedding of the logging statement $l$,  and its corresponding code snippet $s$. Here are several benefits of doing so. First, explicitly identifying the target logging statements allows the model to focus more on the semantic meaning related to target logging statements. Second, it addresses the issue of multiple logging statements within a single method, as specifying $l$ informs the model which statement to analyze.
The loss function, shown in Equation~\ref{equ_loss}, includes a cross-entropy loss and a similarity-based term. The first part is the standard cross-entropy loss~\cite{rumelhart1986learning}, commonly applied in machine learning and optimization areas. The second part is a similarity-based loss term, which forces the target logging statement vectors to be similar to the surrounding code snippet vectors by maximizing their cosine similarity. 
$\alpha$ is a weighting parameter to control the influence of each term. The last term is added to ensure the total loss function remains above zero. 
\begin{equation}
\label{equ_loss}
loss = -(\frac{1}{N}\sum_{i=1}^{N}\sum_{c=1}^{C} y_{i,c} \log(p_{i,c}) +\alpha\frac{1}{N}\sum_{i=1}^{N}\frac{l_i \cdot s_i}{\|l_i\| \|s_i\|} - \alpha) 
\end{equation}

In the equation, \( C \),  \( N \) represent the total number of defect types and samples, respectively. The variable \( y_{i,c} \) is an indicator for the ground truth label for the \( c \)-th type of the \( i \)-th sample, while \( p_{i,c} \) donates the predicted probability for the same type. The vector \( l_i \) and \( s_i \) correspond to the vector representations of the target logging statement and the surrounding code snippet for the \( i \)-th sample, respectively.

\subsection{Updating Detected Defective Logging Statement}
\begin{figure}[t]
    \centering
    \includegraphics[width=0.8\columnwidth]{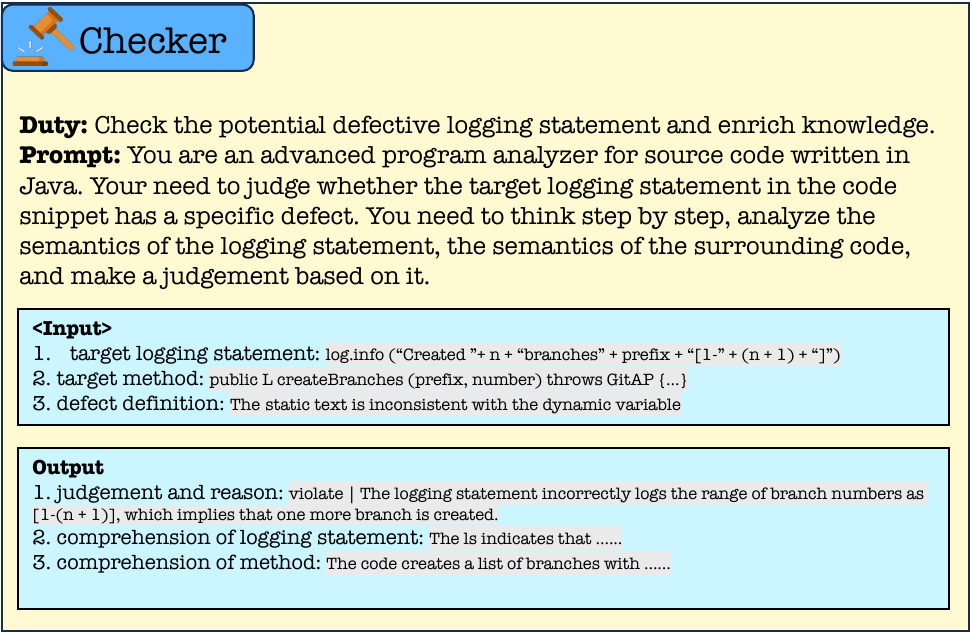}
    \caption{The Checker framework on a static-dynamic inconsistency case.}
    \label{fig:prompt1}
\end{figure}

\begin{figure}[t]
    \centering
    \includegraphics[width=0.8\columnwidth]{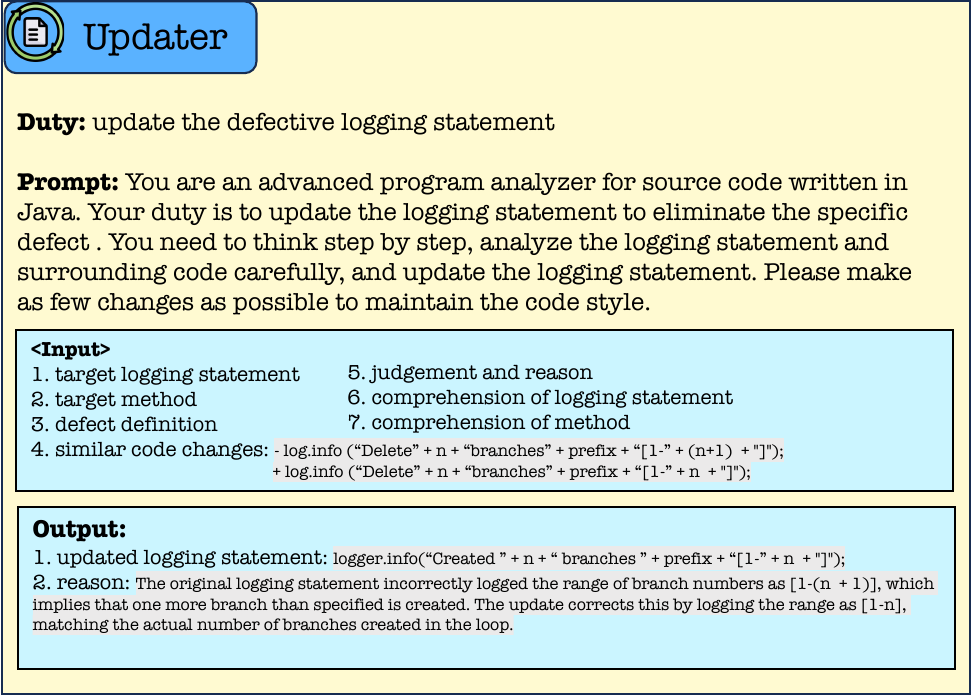}
    \caption{The Updater framework on a static-dynamic inconsistency case.}
    \label{fig:prompt2}
\end{figure}

Following the defect classification phase, where potential defective logging statements are identified, we proceed with 
a multi-expert collaboration framework to repair these statements, inspired by the cognitive synergy~\cite{goertzel2009cognitive}.
In particular, we employ two LLM-based experts: a \textit{checker} and an \textit{updater}.
In Fig.~\ref{fig:prompt1} and Fig.~\ref{fig:prompt2}, we see a representation of the collaborative dynamics between two domain experts. These figures illustrate not only the specialized responsibilities assigned to each expert but also the corresponding prompts tailored for them. Our approach to designing prompts for the updating component is meticulously informed by well-established best practices in LLM prompting. This involves a thoughtful task-specific structuring and the application of few-shot learning, utilizing historical examples which have demonstrated enhancements in task performance in existing literature~\cite{gao2024search,wei2022chain}. Initially, the \textit{checker} plays a crucial role by verifying the classification outcomes generated by the detector. Subsequently, it employs LLMs to augment these results with additional semantic insights. Following this, the role of the \textit{updater} is to diligently execute updates to logging statements in accordance with the semantic guidance provided by the \textit{checker}.

\subsubsection{Checker} 

The role of \textit{checker} is to reduce the false positives brought by the similarity-based detector, where non-defective logging statements could be mistakenly marked as requiring updates due to the extremely imbalanced data in real-world situations. To address this issue, we have designed a \textit{checker} with two objectives. The first objective is to determine whether the detector’s decisions on updating logging statements are necessary, aiming to eliminate false positives. The second objective is to leverage the extensive knowledge of the LLM to offer additional semantic insights, thereby enhancing the effectiveness of subsequent experts.

Specifically, we use the information generated by the detector (i.e., target logging statement, surrounding code, defect type) as the input of the \textit{checker}. Ultimately, the \textit{checker} is expected to generate its judgment, rationale, and semantic information of the logging statements and the surrounding code. This information will be passed on to the next expert for reference.

\subsubsection{Updater}

Upon checking, we identified the target logging statements and their corresponding defect types. The next step involves employing the \textit{updater} to modify these logging statements.

We utilize log-centric changes as supplementary knowledge to guide the LLM in learning the past update patterns from maintainers. We apply a defect type-aware strategy to select proper historical LCC examples for the \textit{updater}’s reference.
In specific, for defect types sensitive to project characteristics (e.g., temporal relation inconsistency and readability issues), we exclusively select the LCCs from the same project to maintain the consistency in the project's logging style~\cite{li2024go}. 
In contrast, for defects related to semantic information inconsistency that are unaffected by logging style and inter-project variations, we incorporate LCCs from multiple projects for the \textit{updater}.
We employ the BM25~\cite{robertson2009probabilistic} ranking algorithm to identify the LCC that are the most similar to the target logging statements. 

\section{Implementation}
\label{implementation}

\subsection{Fine-tuning Dataset}
\label{wellmaintained}
\label{wmr} \label{tdd}

To develop a high-quality dataset for the detector fine-tuning process, our initial step involves the random selection of 20,000 cases that include logging statements, subsequently dividing them into four distinct categories. As a second step, the method detailed in Section~\ref{generate_data} is utilized to produce 5,000 defective instances for each categorized type.
Logging statements sourced from repositories that are meticulously maintained are assumed to be devoid of defects due to their compliance with rigorous community standards and comprehensive maintenance practices.
Specifically, we select \textbf{well-maintained repositories} from the target repositories based on the following criteria:
\begin{itemize}[leftmargin=*]
    \item The initiators of the repository must be companies, such as those from Google, Apache, Facebook, etc. This ensures that the projects are backed by experienced teams leading to better development practices and higher-quality code.
    \item The repository must allow for the submission of issues; open community discussions are beneficial for developers to maintain the project.
    \item Projects must have a license, ensuring the availability of the code for widespread use.
\end{itemize}
We ultimately considered a total of 161 projects, encompassing 128K methods with logging statements, as well-maintained repositories. In addition to 20K defect instances, we also randomly selected 20K examples from the well-maintained repositories as positive samples, leading to 40K samples in total.
We split the data into training, validation, and test sets with a ratio of 8:1:1 using stratified sampling~\cite{pirzadeh2011concept}. This results in a total of 32K samples for fine-tuning the detector.

\subsection{Implementation Details}

During LLM-based defective logging statement synthesis, we use the GPT-4o~\cite{OpenAIGPT4o} API provided by Openai as the default LLM model. We use UniXcoder-base as the tokenization and embedding tool. We use Adam~\cite{kingma2017adammethodstochasticoptimization} as the optimizer and set the learning rate to 5e-5, Adam epsilon to 1e-8, and dropout rate to 0.1 when fine-tuning the similarity-based classification model. We fine-tune the model for 10 epochs and set the max token number to 1024. We run them on a server with Dual Intel(R) Gold 6326 CPU, 512GB physical memory, and
4x NVIDIA RTX 3090 GPU. The OS version is Ubuntu 18.04. For all experiments related to LLM inference, we set the temperature to 0 so that LLM would generate the same output for the same query to ensure reproducibility.
\section{Evaluation}
\label{evaluation}

We evaluate \name by answering the following research questions:

\textbf{RQ1: How effective is \name at detecting and enhancing defective logging statements?} 

\textbf{RQ2: How effective is \name at end-to-end detect and update defective logging statements?}

\textbf{RQ3: How do different components affect the performance of \name?}

\textbf{RQ4: How generalizable is \name for different LLM backbones?}

\textbf{RQ5: How is the practitioners' feedback on \name?}

\subsection{Evaluation Dataset}
To evaluate the performance of \name, we conducted experiments on three distinct datasets.
\subsubsection{Existing logpatches}
\label{logpatch}
In our study, we employ the dataset annotated during the manual investigation presented in (\ref{manualstudy}) as a source of existing logpatches. These consist of 258 defective instances derived from the manually categorized dataset. Furthermore, we enhance our dataset by sampling an additional 258 logging statements that are defect-free from repositories known for their meticulous maintenance as outlined in Section~\ref{wellmaintained}. This procedure results in a comprehensive dataset comprising a total of 516 cases. It is important to highlight that these specific data points are deliberately excluded from the training dataset.
\subsubsection{Synthetic test data}
As mentioned by~\ref{tdd}, we used a stratified sampled test set including 4,000 samples, as the synthetic test data. 

\subsubsection{New repositories}
\label{new_repositories}
To evaluate the effectiveness of \name in addressing previously unidentified defects, we randomly sampled 100K log-code pairs from the target repositories but outside the collected well-maintained dataset for human evaluation. These samples were deliberately excluded from all previous processes, thereby guaranteeing that the evaluation was conducted on entirely fresh data.

\subsection{Metrics}
We evaluate \name from two perspectives: the detection ability and updating ability.

\subsubsection{Evaluation metrics for detection ability}
For the detection ability of \name, we use Precision, Recall, and F1-macro to measure multiple classification performance.
Formally, $Precision = \frac{TP}{TP+FP}$, $Recall = \frac{TP}{TP+FN}$, $F_1 = 2*(\frac{Precision*Recall}{Precision+Recall})$, and $F_{1}^{macro}=\frac{1}{N} \sum_{i=1}^{N}F_{1}^i$. We use $F_{1}^{macro}$ to give equal weight to each class, avoiding the issue of inflating the weight of the class with more instances. 

\subsubsection{Evaluation metrics for updating ability}
For the updating ability, we use similar metrics as logging statement generation task. Different metrics are used to evaluate different parts of the logging statements.
    
(1) Static content. In alignment with recent research efforts~\cite{li2023exploring,mastropaoloUsingDeepLearning2022,dingLoGenTextAutomaticallyGenerating2022}, the quality of generated logging texts is evaluated using two established metrics from the field of machine translation: BLEU~\cite{papineni2002bleu} and ROUGE~\cite{lin2004rouge}. These n-gram-based metrics quantify the similarity between the generated log messages and those crafted by developers, producing a normalized score ranging from 0 to 1, where higher scores denote greater textual correspondence. Specifically, we utilize variations of these metrics, namely BLEU-K (where K = {1, 2, 4}) and ROUGE-K (where K = {1, 2, L}), to measure the overlap in terms of K-grams between the generated and actual log texts.

(2) Dynamic variables. Following the recent work~\cite{li2023exploring}, we utilize special Precision, Recall, and F1 to assess the accuracy of updated logging variables
. For each logging statement, let $S_{ud}$ represent the variables in the \name updates, and $S_{gt}$ denote the variables in the actual logging statement. We calculate and report the proportion of correctly updated variables ($precision$=$\frac{S_{ud} \cap S_{gt}}{S_{ud}}$), the proportion of actual variables that are correctly predicted by the model ($recall$=$\frac{S_{ud} \cap S_{gt}}{S_{gt}}$), and their harmonic mean ($F_1$=$2*\frac{Precision*Recall}{Precision+Recall}$).

The above metrics of updating evaluation, are widely used to assess the quality of text generation by calculating the similarity between generated log text and actual ones. However, a high similarity score in this task is not necessarily related to a significant improvement, as the defective logging statements prior to the update could already bear a close resemblance to the ground truth (i.e., the logging statements updated by developers).
\textbf{Consequently, we propose a novel metric named Improvement Coefficient (IC), which focuses on the change in these metrics before and after the updates.} The formula~\ref{kappa} reflects the proportion of the improvement achieved after the updates relative to the maximum possible improvement.
Specifically, $m \in \{BLEU, ROUGE, Precision, Recall, F_1\}$. $m_{origin}$ and $m_{update}$ denote the metrics before and after the updates, respectively.
\begin{equation}
     \text{Improvement Coefficient (IC)} = \frac{m_{updated}-m_{origin}}{1-m_{origin}}
\label{kappa}
\end{equation}

\begin{table}[tbp]
\centering
\caption{The detection result on existing logpatches and synthetic data}
\label{tab:rq1_res_detection}
\begin{tabular}{@{}lcccccc@{}}
\toprule
                  & \multicolumn{3}{c}{Existing logpatches} & \multicolumn{3}{c}{Synthetic data} \\ \midrule
                  & Precision      & Recall     & F1        & Precision    & Recall    & F1      \\ \midrule
CodeT5+           & 0.314          & 0.062      & 0.096     & 0.366        & 0.075     & 0.115   \\
Claude3.5-sonnet  & 0.441               &  0.436          &   0.413        & 0.575             &  0.492         &  0.530       \\
Deepseek-coder-v2 & 0.406          & 0.345      & 0.351     & 0.648             &  0.479         &  0.512       \\
Deepseek-R1       & 0.381          & 0.426      & 0.391     & 0.419        & 0.564     & 0.446   \\
GPT3.5            & 0.256          & 0.204      & 0.167     & 0.474        & 0.258     & 0.270   \\
GPT4o             & 0.468          & 0.401      & 0.410     & 0.512        & 0.467     & 0.476   \\
\name        & 0.725          & 0.552      & 0.625     & 0.925        & 0.673     & 0.776  \\
\bottomrule
\end{tabular}
\end{table}

\subsection{RQ1: Effectiveness of \name}

\begin{table}[t]
\centering
\caption{The static texts updating results on existing logpatches and synthetic data }
\label{tab:static_text}
\resizebox{\textwidth}{!}{%
\begin{tabular}{lllllll}
\toprule
                        & BLEU-1                      & BLEU-2                      & BLEU-4                      & ROUGE-1                     & ROUGE-2                     & ROUGE-L                     \\ \midrule
\rowcolor{grey}\multicolumn{7}{c}{Existing logpatches}                                                                                                                                                                     \\
\textbf{GPT4o-Instruct} & 0.690 (4.90\%$\uparrow$)    & 0.587 (12.86\%$\uparrow$)   & 0.268 (0.94\%$\uparrow$)    & 0.746 (13.01\%$\uparrow$)   & 0.593 (17.44\%$\uparrow$)   & 0.745 (12.96\%$\uparrow$)   \\
\textbf{GPT4o-ICL}      & 0.681 (2.14\%$\uparrow$)    & 0.588 (13.08\%$\uparrow$)   & 0.262 (0.14\%$\uparrow$)    & 0.750 (14.38\%$\uparrow$)   & 0.603 (19.47\%$\uparrow$)   & 0.748 (13.99\%$\uparrow$)   \\
\textbf{GPT4o-DET}      & 0.701 (8.28\%$\uparrow$)    & 0.620 (19.83\%$\uparrow$)   & 0.327 (8.93\%$\uparrow$)    & 0.781 (25.00\%$\uparrow$)   & 0.650 (29.01\%$\uparrow$)   & 0.779 (24.57\%$\uparrow$)   \\
\textbf{\name}                   & 0.799 (38.34\%$\uparrow$)   & 0.721 (41.13\%$\uparrow$)   & 0.384 (16.64\%$\uparrow$)   & 0.849 (48.28\%$\uparrow$)   & 0.738 (46.85\%$\uparrow$)   & 0.848 (48.12\%$\uparrow$)   \\ \midrule
\rowcolor{grey}\multicolumn{7}{c}{Synthetic data}                                                                                                                                                                          \\
\textbf{GPT4o-Instruct} & 0.546 (-7.32\%$\downarrow$) & 0.465 (-7.86\%$\downarrow$) & 0.316 (-2.54\%$\downarrow$) & 0.634 (-6.08\%$\downarrow$) & 0.499 (-9.15\%$\downarrow$) & 0.632 (-6.35\%$\downarrow$) \\
\textbf{GPT4o-ICL}      & 0.545 (-7.56\%$\downarrow$) & 0.468 (-7.25\%$\downarrow$) & 0.313 (-3.00\%$\downarrow$) & 0.639 (-4.64\%$\downarrow$) & 0.508 (-7.19\%$\downarrow$) & 0.637 (-4.91\%$\downarrow$) \\
\textbf{GPT4o-DET}      & 0.640 (14.89\%$\uparrow$)   & 0.579 (15.12\%$\uparrow$)   & 0.371 (5.69\%$\uparrow$)    & 0.715 (17.39\%$\uparrow$)   & 0.636 (20.69\%$\uparrow$)   & 0.714 (17.34\%$\uparrow$)   \\
\textbf{\name}                   & 0.659 (19.39\%$\uparrow$)   & 0.592 (17.74\%$\uparrow$)   & 0.436 (15.44\%$\uparrow$)   & 0.753 (28.40\%$\uparrow$)   & 0.648 (23.31\%$\uparrow$)   & 0.751 (28.03\%$\uparrow$)   \\ \bottomrule
\end{tabular}%
}
\end{table}

\begin{table}[t]
\centering
\caption{The dynamic variables updating result on existing logpatches and synthetic data}
\small
\label{tab:dynamic_varialbe}
\resizebox{0.75\textwidth}{!}{
\begin{tabular}{llll}
\toprule
                        & Precision                 & Recall                    & F1                        \\ \midrule
\rowcolor{grey}\multicolumn{4}{c}{Existing logpatches}                                                                     \\
\textbf{GPT4o-Instruct} & 0.272 (0.13\%$\uparrow$)  & 0.256 (1.59\%$\uparrow$)  & 0.264 (1.15\%$\uparrow$)  \\
\textbf{GPT4o-ICL}      & 0.300 (3.97\%$\uparrow$)  & 0.307 (21.82\%$\uparrow$) & 0.304 (16.47\%$\uparrow$) \\
\textbf{GPT4o-DET}      & 0.315 (6.03\%$\uparrow$)  & 0.296 (17.46\%$\uparrow$) & 0.305 (16.85\%$\uparrow$) \\
\textbf{\name}          & 0.439 (23.04\%$\uparrow$) & 0.451 (26.60\%$\uparrow$) & 0.445 (24.90\%$\uparrow$) \\ \midrule
\rowcolor{grey}\multicolumn{4}{c}{Synthetic data}                                                                          \\
\textbf{GPT4o-Instruct} & 0.696 (14.85\%$\uparrow$) & 0.707 (42.20\%$\uparrow$) & 0.702 (32.57\%$\uparrow$) \\
\textbf{GPT4o-ICL}      & 0.683 (11.20\%$\uparrow$) & 0.664 (33.73\%$\uparrow$) & 0.673 (26.02\%$\uparrow$) \\
\textbf{GPT4o-DET}      & 0.752 (30.53\%$\uparrow$) & 0.761 (52.85\%$\uparrow$) & 0.756 (44.79\%$\uparrow$) \\
\textbf{\name}          & 0.795 (42.57\%$\uparrow$) & 0.778 (56.21\%$\uparrow$) & 0.787 (51.80\%$\uparrow$) \\ \bottomrule
\end{tabular}
}
\end{table}

\textbf{Approach.}
The effectiveness of \name should be evaluated from two aspects: first, the detection capability, specifically the capability to identify defective logging statements;  second, the updating capability, which pertains to the similarity between the revised logging statements (i.e., prediction) and the actual ground truth (i.e., repository update history). 

To evaluate detection capabilities, we employ a \textbf{CodeT5+}~\cite{wang2023codet5+} model pre-trained following the CMI-Finder~\cite{bouzenia2023When} methodology as a baseline for comparison. This model has demonstrated effective performance in binary classification tasks concerning message-condition inconsistency. We adhered to the same pretraining methodology as our proposed detector, utilizing an identical fine-tuning dataset consisting of 32K samples, conducted fine-tuning over 10 epochs, and capped the maximum token length at 1024. Furthermore, given the progress in the development of LLM and their potential use in similar tasks, we also consider serval prominent LLMs as baselines, including gerneral-purpose LLMs (Claude3.5-sonnet, GPT3.5, GPT4o), code-based LLMs (Deepseek-coder-v2), and reasoning LLMs (Deepseek-R1). For the LLM-based baselines, we provide five fixed examples for task demonstration. The prompt we use for the baselines can be found in~\cite{comparisonbaselineprompt}.

For the updating ability, due to the lack of related research, we employed several different prompting methods as baselines. Specifically, we utilized instruction prompting \textbf{(Instruct)}, in-context learning (\textbf{ICL}), and prompts with detection information (\textbf{DET}). Instruction prompting involves directly instructing the LLM to update defective logging statements without providing additional context. In ICL, we retrieve examples to help the model better understand the objective; we randomly sampled an add-delete pair from the LCC to create these examples. For DET, we provided the LLM relevant information about the type of defect and used prompts to guide the LLM to first explain why the target logging statement contains the specific defect before updating it.

\textbf{Result.}
Table~\ref{tab:rq1_res_detection} shows the detection results with different classification models. 
The experimental results demonstrate that \name significantly outperforms all baseline models in detecting defective logging statements across both existing logpatches and synthetic datasets. With F1 scores of 0.625 and 0.776 respectively, \name achieves substantial improvements over the next best performers Claude3.5-sonnet. This superior performance is particularly evident in precision metrics, where \name reaches 0.725 and 0.925 across the two datasets, indicating its high reliability in correctly identifying defective logging statements. General-purpose LLMs like Claude3.5-sonnet and GPT4o show moderate effectiveness (F1 scores around 0.41-0.47), while code-specialized models like Deepseek-coder-v2 perform comparably on synthetic data but lag on real-world logpatches. Furthermore, all models perform better on synthetic data than on existing logpatches, suggesting that real-world logging defects present more complex patterns than synthetically generated ones. These findings validate the effectiveness of our specialized approach for detecting logging inconsistencies in production code.

Table~\ref{tab:static_text} and Table~\ref{tab:dynamic_varialbe} illustrate the updating performance on static content and dynamic variables, respectively. 
Compared to the three baselines, \name demonstrates significant performance improvements across various metrics on two datasets. Specifically, for GPT4o-Instruct and GPT4o-ICL, which lack defect information, the improvement on the logpatches dataset is minimal (0.14\%$\uparrow$ IC on BLEU-4), and there is even a regression (3.00\%$\downarrow$) on synthetic data. However, GPT4o-DET, which includes defect information, shows a certain level of improvement, highlighting the importance of defect information.

\begin{tcolorbox}[boxsep=1pt,left=2pt,right=2pt,top=3pt,bottom=2pt,width=\columnwidth,colback=white!95!black,boxrule=1pt, colbacktitle=white!30!black,toptitle=2pt,bottomtitle=1pt,opacitybacktitle=0.4]
\textbf{Answer to RQ1:} \name demonstrates strong capabilities in defect detection and updating. It outperforms all baselines in detection with an F1 score of 0.625. Regarding its update ability, it improves static text and dynamic variables by 48.12\% and 24.90\%, respectively.
\end{tcolorbox}

\subsection{RQ2: The end-to-end effectiveness of \name}
\textbf{Approach.} In this RQ, our goal is to explore the efficacy of \name once it integrates the tasks of detecting and updating defective logging statement. We conduct a thorough evaluation and comparison between \name and several leading LLMs: a general-purpose model (GPT4o), a code-based model (Deepseek-coder-v2), and a reasoning model (Deepseek-R1). For each of these LLM-based benchmark models, we supply five examples for ICL to facilitate task demonstration. Details of the prompt used for the end-to-end task are available in ~\cite{comparisonbaselineprompt}. We assess these models using the existing logpatches dataset, as elaborated in Section~\ref{logpatch}, focusing on metrics such as Detection Precision, Recall, F1-Score, ROUGE-L, and Variable-F1. It should be noted that for each model, the computation of updating ability metrics is limited to cases in which a defective detection is successfully achieved. Additionally, to appraise the financial cost associated with each method, we introduce the concept of cost per input (CPI), representing the average cost of using the official API of LLM for each input that aims at the detection and revision of defective logging statements. Note that the base LLM model for LogUpdater is GPT4o.

\textbf{Result.} Table~\ref{tab:end2end} presents the evaluation results for end-to-end comparisons of \name with the baselines. We can see the \name outperforms all baselines in terms of all metrics. \name demonstrates significant performance enhancements compared to all baseline models when evaluated using the metrics of Precision, Recall, and F1-score. Specifically, \name achieves a precision of 0.725 and recall of 0.552, resulting in an F1-score of 0.625, which significantly surpasses the baseline models (0.373, 0.372, and 0.337 for GPT4o, Deepseek-coder-v2, and Deepseek-R1, respectively). That means, for the detection performance, \name detects more defectifve logging statements with higher precision.

In terms of the updating metrics, we observe that \name demonstrates competitive performance in ROUGE-L (0.928) and Variable-F1 (0.977) metrics. Notably, \name shows substantial improvements over original logging statements as indicated by the percentage increases: 81.29\% in ROUGE-L and 55.77\% in Variable-F1. This suggests that \name is particularly effective at fixing the defective logging statements at both the static content aspect and dynamic variable aspect.

From an efficiency perspective, \name demonstrates remarkable cost-effectiveness compared to its base model GPT4o. With a CPI of 0.0011, \name is approximately 16.2 times more cost-efficient than the direct use of GPT4o (0.0194). This substantial cost reduction while maintaining superior performance is particularly noteworthy since both approaches utilize GPT4o as their foundation. The significant cost advantage of \name can be attributed to its specialized architecture (i.e., use small model to develop a detector), which minimizes the number of API calls required for each input cases analysis. This cost-effectiveness, combined with \name's superior detection and generation performance, makes it a more viable solution for large-scale logging statement maintenance in real-world applications.

\begin{table}[tbp]
\centering
\caption{The end-to-end comparisons of \name with the baselines.}
\label{tab:end2end}
\begin{tabular}{@{}lccccccc@{}}
\toprule
                  & Precision & Recall & F1-score & ROUGE-L & Variable-F1 & CPI\\ \midrule
GPT4o             & 0.401          & 0.403       & 0.373           &  0.905 (68.33\%$\uparrow$)       &  0.967 (29.79\%$\uparrow$)        & 0.0194   \\
Deepseek-coder-v2 & 0.383          & 0.365       & 0.372            &  0.861 (51.90\%$\uparrow$)       &  0.979 (0.00\%)      &0.0022    \\
Deepseek-R1       & 0.358          & 0.414       & 0.337            &  0.897 (62.68\%$\uparrow$)       &  0.905 (12.03\%$\uparrow$)   &0.0044        \\
\name        & 0.725          & 0.552       & 0.625            &  0.928 (81.29\%$\uparrow$)       &   0.977 (55.77\%$\uparrow$)  & 0.0011        \\ \bottomrule
\end{tabular}
\end{table}

\begin{tcolorbox}[boxsep=1pt,left=2pt,right=2pt,top=3pt,bottom=2pt,width=\columnwidth,colback=white!95!black,boxrule=1pt, colbacktitle=white!30!black,toptitle=2pt,bottomtitle=1pt,opacitybacktitle=0.4]

\textbf{Answer to RQ2:} In summary, \name significantly outperforms baseline models in terms of all metrics by significant margins. Moreover, \name is 16.2 times more cost-efficient than GPT4o, enhancing its viability for practical applications.

\end{tcolorbox}

\subsection{RQ3: Effects of Different Components}
\textbf{Approach.} 
We consider the contribution of different components to \name’s performance from two perspectives. 

For detection capability, we first experiment with several different configurations on PLM to understand how hyperparameters affect our detection results. Specifically, we change the epoch in fine-tuning and the $\alpha$ in loss function to control the importance of traditional cross-entropy loss and cosine similarity regularization term.
Then we remove the checker LLM agent and use only the PLM for detection as a comparison. This will demonstrate the necessity of using the checker. 

To assess the updating capability, we begin by comparing the outcomes with a scenario in which the detection module is completely omitted, allowing the LLM to directly update the logging statements requiring enhancement. This approach will highlight the importance of including the detection module. Subsequently, we conduct an analysis to evaluate how varying the number of similar LCCs presented in the prompt influences the effectiveness of the Updater.

\begin{figure}[tbp]
    \centering
    \begin{minipage}{0.4\textwidth}
        \centering
        \includegraphics[height=4cm]{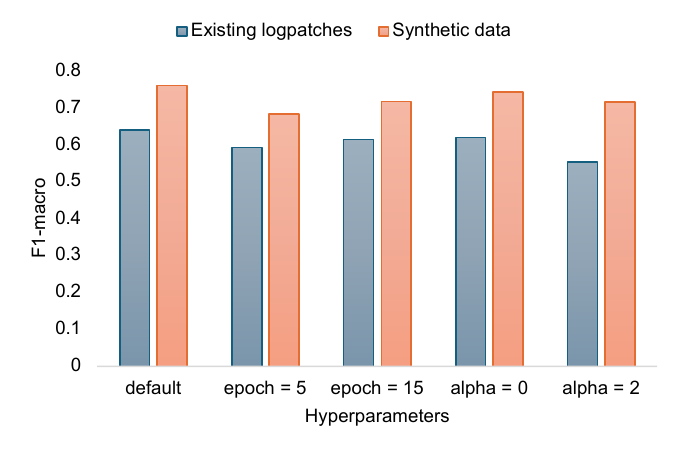}
        \caption{Hyperparameter analysis.}
        \label{fig:sub1}
    \end{minipage}
    \hfill
    \begin{minipage}{0.45\textwidth}
        \centering
        \includegraphics[height=4cm]{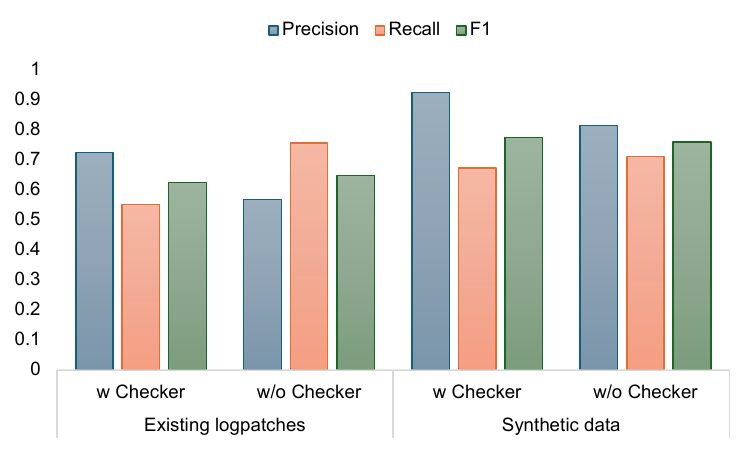}
        \caption{Detection results w/wo checker.}
        \label{fig:sub2}
    \end{minipage}
\end{figure}

\textbf{Result.}
For hyperparameters in PLM, the default is ten epochs and $\alpha$ is 0.5. Experimental results in figure~\ref{fig:sub1} indicate that both reducing the training by 5 epochs (0.05$\downarrow|$ 0.08$\downarrow$) or extending it by 5 epochs (0.03$\downarrow|$ 0.04$\downarrow$) lead to a certain degree of performance degradation on both existing logpatches and synthetic data. Regarding the hyperparameter $\alpha$, setting it to 0 (using BCE loss directly) or setting it to 2 (giving higher weight to the similarity-based term) also results in significant performance degradation. This demonstrates the effectiveness of our designed loss function.

Figure~\ref{fig:sub2} illustrates the detection performance of the framework with and without the checker. We observe that using the checker does not result in a significant change in the F1 score. Additionally, the precision on both datasets is significantly improved (0.15$\uparrow|$ 0.11$\uparrow$), while the recall is notably reduced (0.20$\downarrow|$ 0.04$\downarrow$). This aligns with our design rationale for the checker, which is to filter out the massive false positives generated by the similarity-based classifier.

\begin{figure}[tbp]
    \centering
    \includegraphics[height=4.5cm]{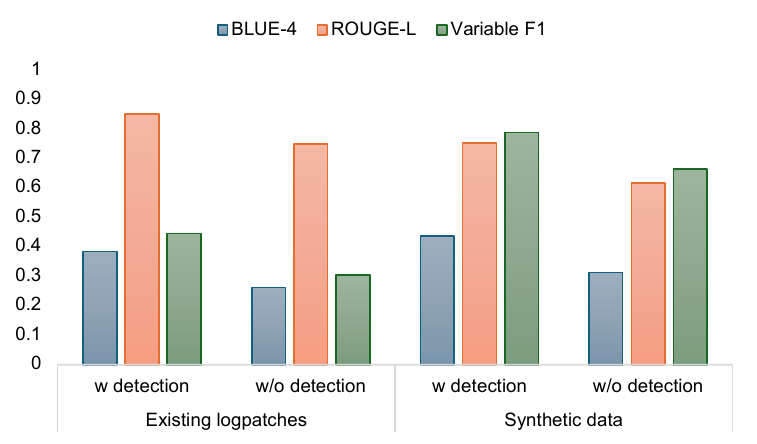}
    \caption{Updating results w/wo detection.}
    \label{fig:sub3}
\end{figure}

Figure~\ref{fig:sub3} illustrates the update performance of the framework with and without the detection component. We observe a significant decrease in various evaluation metrics across both datasets when the detection component is not used. 
Additionally, even without the detection component, the update component still provides some degree of repair (13.99\%$\uparrow$ IC on ROUGE-L) on the existing logpatches. However, its performance on the synthetic dataset is worse than not performing updates at all (10.98\%$\downarrow$ IC on ROUGE-L).

Table~\ref{tab:number_analysis} demonstrates the performance during updating across varying numbers of similar LCCs as outlined in the prompt. It is evident that all performance metrics show a gradual increase when the number of shots is incremented from 1 to 4, for the case of existing log patches, and up to 3 for synthetic data. This trend reaches its highest point at 4 shots, recording a peak value of 0.395 for BLEU-4, which signifies a 2.8\% increase compared to the 1-shot scenario. Such findings indicate that the updater gains from a more substantial provision of examples in the prompt, thereby facilitating an improved contextual comprehension and consequently enhancing the preciseness of the logging statements. Nonetheless, beyond this optimal range, when the number of shots is further escalated to 5 and 10, a slight decrease is observed across all metrics.

\begin{table}[tbp]
\centering
\caption{Shot Number Analysis}
\label{tab:number_analysis}
\begin{tabular}{@{}lllllll@{}}
\toprule
\multicolumn{1}{c}{Shot Number} & \multicolumn{1}{c}{1} & \multicolumn{1}{c}{2} & \multicolumn{1}{c}{3} & \multicolumn{1}{c}{4} & \multicolumn{1}{c}{5} & \multicolumn{1}{c}{10} \\ \midrule
\rowcolor{grey}\multicolumn{7}{c}{Existing logpatches}                                                                                                                                          \\
BLEU-4                          & 0.384                 & 0.387                 & 0.388                 & \textbf{0.395}                 & 0.391                 & 0.385                  \\
ROUGE-L                         & 0.848                 & 0.850                 & 0.855                 & \textbf{0.864}                 & 0.858                 & 0.844                  \\
Variable-F1                     & 0.445                 & 0.438                 & 0.447                 & \textbf{0.462}                 & 0.449                 & 0.434                  \\
\rowcolor{grey}\multicolumn{7}{c}{Synthetic data}                                                                                                                                               \\
BLEU-4                          & 0.436                 & 0.440                 & \textbf{0.446}                 & 0.443                 & 0.442                 & 0.438                  \\
ROUGE-L                         & 0.751                 & 0.757                 & \textbf{0.762}                 & 0.759                 & 0.760                  & 0.754                  \\
Variable-F1                     & 0.787                 & 0.792                 & 0.792                 & \textbf{0.796}                 & 0.789                 & 0.784                  \\ \bottomrule
\end{tabular}
\end{table}

\begin{tcolorbox}[boxsep=1pt,left=2pt,right=2pt,top=3pt,bottom=2pt,width=\columnwidth,colback=white!95!black,boxrule=1pt, colbacktitle=white!30!black,toptitle=2pt,bottomtitle=1pt,opacitybacktitle=0.4]

\textbf{Answer to RQ3:} Each component of \name contributes variously to its overall performance. Removing the detection module notably degrades performance, underscoring its critical role in enhancing logging statement accuracy. 4-shot LCCs in prompt maximizes the update performance of \name.

\end{tcolorbox}
\subsection{RQ4: Generalizability of \name}
\textbf{Approach.}
We evaluate the capability of \name using code-specific (Phind-CodeLlama~\cite{PhindCodeLlama34Bv2}, Deepseek-coder~\cite{zhu2024deepseek}) and general (Claude3.5-sonnet~\cite{anthropic2024claude}, Llama3~\cite{llama3modelcard}, GPT3.5~\cite{gpt-3.5}, GPT4o~\cite{OpenAIGPT4o}) LLMs on both synthetic data and existing logpatches. This can illustrate the generalizability of our framework, which means that we don't rely on the ability of specific LLM backbones.

\textbf{Result.}
Table~\ref{tab:backbone} shows the performance of \name with six different backbone models. First, we observe that our framework can consistently perform well on most backbones in terms of all metrics by a large margin. On average, all models get 0.614 F1-macro for detection ability. From the perspective of update ability, models have an average of IC 13.19\%$\uparrow$ (reflected by BLEU-4), 38.68\%$\uparrow$ (reflected by ROUGE-L), and 12.52\%$\uparrow$ (reflected by F1) respectively.

The results demonstrate the generalizability of \name across various backbone models. We anticipate that the performance of \name can be further enhanced with advancements in LLMs.

\begin{table}[t]
\centering
\caption{The performance of \name with different backbone models.}
\small
\resizebox{0.8\textwidth}{!}{
\label{tab:backbone}
\begin{tabular}{lcccc}
\toprule
                 & Detection & \multicolumn{2}{c}{Updating text} & Updating variable \\ \midrule
                 & F1-macro  & BLEU-4          & ROUGE-L         & F1                \\ \toprule
CodeLlama-34b-v2        &  0.613     & 0.377 (15.70\%$\uparrow$)          & \textbf{0.856 (50.85\%$\uparrow$)}          & 0.319 (7.85\%$\uparrow$)            \\
Deepseek-coder-v2   &  \textbf{0.642}     & 0.371 (14.88\%$\uparrow$)          & 0.818 (37.88\%$\uparrow$)          & 0.309 (6.50\%$\uparrow$)            \\
Claude3.5-sonnet &  0.637     & 0.355 (12.72\%$\uparrow$)          & 0.810 (36.15\%$\uparrow$)          & 0.359 (13.26\%$\uparrow$)            \\
Llama3-70b           &  0.594     & 0.353 (12.45\%$\uparrow$)           & 0.819 (38.23\%$\uparrow$)           & 0.359 (13.26\%$\uparrow$)            \\
GPT3.5           &  0.573     & 0.311 (6.77\%$\uparrow$)        & 0.771 (21.84\%$\uparrow$)        & 0.330 (9.34\%$\uparrow$)            \\
GPT4o (original) &  0.625     & \textbf{0.384 (16.64\%$\uparrow$)}          & 0.848 (48.12\%$\uparrow$)          & \textbf{0.445 (24.90\%$\uparrow$)}            \\ \bottomrule
\end{tabular}%
}
\end{table}

\begin{tcolorbox}[boxsep=1pt,left=2pt,right=2pt,top=3pt,bottom=2pt,width=\columnwidth,colback=white!95!black,boxrule=1pt, colbacktitle=white!30!black,toptitle=2pt,bottomtitle=1pt,opacitybacktitle=0.4]

\textbf{Answer to RQ4:} \name showcases great generalization ability by consistently enhancing the logging statements across various LLMs, including code-specific and general ones.

\end{tcolorbox}
\subsection{RQ5: Practitioners' feedback on \name}

In this RQ, we evaluate \name based on the practitioners' feedback from two perspectives. 

\textit{1) Human evaluation}

\textbf{Approach.}
We aim to assess if \name can accurately identify and update unseen defective logging statements while achieving a reasonable successful rate (SR). For this, we employ the new repositories dataset (Sec. ~\ref{new_repositories}), which has \textit{no prior} identified logging statement defects and modifications. Two authors manually verified each case to confirm whether the detection is accurate and if the update is correct. If both annotators unanimously agree, we consider it a success case. Finally, we compute the success rate as the ratio of the number of successful cases to the total number of cases.

\textbf{Result.} \name reports 149 defective logging statements out of 100K data.
The two authors achieve a Cohen's Kappa of 0.84 and finally identify 91 successful detections and updates after reaching a consensus, resulting in a 61.49\% SR. 
This outcome demonstrates the low false positive rate of \name, enabling developers to conserve time in scrutinizing meaningless updates.
Table~\ref{tab:rq4_sr} demonstrates SR for \name across different types of defects, with the highest SR of 88.57\% observed for readability issues, and the lowest at 46.15\% for static-dynamic inconsistency.

\textit{2) Developer's feedback}

\textbf{Approach.}
In order to comprehensively grasp the significance of rectifying identified defects from the viewpoint of developers, we conduct a randomized sampling of \textbf{40 cases} from the previously 91 instances of successfully updated logging cases. We then submit pull requests (PRs) to the repository maintainers for each sampled instance. Each pull request was accompanied by detailed information regarding the types of defects encountered, in addition to the enhanced logging statements suggested by \name. The developers' readiness and commitment to dedicate their efforts towards testing, validating, and ultimately accepting our submitted PRs underscores the effectiveness and practical value of our proposed strategy. The full merged cases are shown in our replication package~\cite{package}.

\textbf{Result.} To date, \textbf{25} out of 40 proposed PRs have been accepted and merged into the codebases, and an additional 3 cases have been recognized as bugs and are incorporated in future versions.

\begin{table}[t]
\caption{The successful rate for \name in new data.}
\centering
\small
\label{tab:rq4_sr}
\begin{tabular}{l|cccc}
\toprule
Defect type     & statement-code incon.    & static-dynamic incon.    & temporal incon.      & readability issues      \\ \toprule
Successful rate & 62.07\% & 46.15\% & 63.16\% & 88.57\% \\ \bottomrule
\end{tabular}

\end{table}

\begin{table}[t]
\centering
\footnotesize
\caption{The real-world defective logging statement update cases suggested by \name.}
\label{tab:rq5}
\begin{tabular}{p{2.8cm}p{9cm}}\toprule
Information & Update code$^\dagger$  \\ \toprule
Statement-code inconsistency \textcolor{codegreen}{\textbf{merged}} by infinispan~\cite{statement_code_2024}           &    \begin{adjustbox}{valign=t}
                \begin{lstlisting}[language=Java, basicstyle=\ttfamily\scriptsize, frame=none, backgroundcolor=\color{white},showstringspaces=false]
public CompletionStage<Void> rollbackAsync (){
-   log.tracef("Transaction.commit() invoked with Xid=%s", xid);
+   log.tracef("Transaction.rollback() invoked with Xid=%s", xid);
    status = Status.STATUS_MARKED_ROLLBACK;
\end{lstlisting}
\end{adjustbox}                          \\ \midrule
Static-dynamic inconsistency \textcolor{codegreen}{\textbf{merged}} by trino~\cite{static_dynamic_2024}           &       \begin{adjustbox}{valign=t}
                \begin{lstlisting}[language=Java, basicstyle=\ttfamily\scriptsize, frame=none, backgroundcolor=\color{white},showstringspaces=false]
 private Comparator internalCompilePageWithPositionComparator(List<Type> types, List<Integer> sortChannels, List<SortOrder> sortOrders){
-   log.error(t, "Error compiling comparator for channels %s with order %s", sortChannels, sortChannels);
+   log.error(t, "Error compiling comparator for channels %s with order %s", sortChannels, sortOrders);
    comparator = new SimplePageWithPositionComparator(types, sortChannels, sortOrders, typeOperators);
\end{lstlisting}
\end{adjustbox}             \\ \midrule
Temporal inconsistency \textcolor{codegreen}{\textbf{merged}} by openhab-addons~\cite{temporal_inconsistency_2024} &  \begin{adjustbox}{valign=t}
                \begin{lstlisting}[language=Java, basicstyle=\ttfamily\scriptsize, frame=none, backgroundcolor=\color{white},showstringspaces=false]
public void run() {
-   logger.debug("Receiver thread started");
+   logger.debug("Starting receiver thread");
    while (!interrupted()) {
        Optional<String> message = readLineBlocking();}
    logger.debug("Receiver thread finished");
\end{lstlisting}
\end{adjustbox}                             \\ \midrule
Readability issue \textcolor{codegreen}{\textbf{merged}} by openhab-addons~\cite{readability_issues_2024}   &
            \begin{adjustbox}{valign=t}
                \begin{lstlisting}[language=Java, basicstyle=\ttfamily\scriptsize, frame=none, backgroundcolor=\color{white},showstringspaces=false]
public void handleCommand(ChannelUID cUID, Command cmd) {
    if (command instanceof RefreshType) {
        logger.debug("Intellflo received refresh command");
        logger.debug("IntelliFlo received refresh command");
        updateChannel(channelUID.getId(), null);}}
\end{lstlisting}
\end{adjustbox}
   \\    \midrule
A \textcolor{Mahogany}{\textbf{false positive case}} offered from \name.  &
            \begin{adjustbox}{valign=t}
                \begin{lstlisting}[language=Java, basicstyle=\ttfamily\scriptsize, frame=none, backgroundcolor=\color{white},showstringspaces=false]
public Future scaleDown(Reconciliation reconciliation, String namespace, String name, int scaleTo, long timeoutMs) {
    Integer nextReplicas = currentScale(namespace, name);
    while (nextReplicas > scaleTo) {
        nextReplicas--;
        LOGGER.infoCr(reconciliation, "Scaling down from {} to {}", nextReplicas + 1, nextReplicas);
        resource(namespace, name).withTimeoutInMillis(timeoutMs).scale(nextReplicas);}
    return nextReplicas;}
\end{lstlisting}
\end{adjustbox}
            \\    \bottomrule
\multicolumn{2}{l}{\parbox{0.8\linewidth}{\scriptsize $^\dagger$ To save space, some parts of the update code snippets are abbreviated.}}
\end{tabular}
\end{table}

\begin{tcolorbox}[boxsep=1pt,left=2pt,right=2pt,top=3pt,bottom=2pt,width=\columnwidth,colback=white!95!black,boxrule=1pt, colbacktitle=white!30!black,toptitle=2pt,bottomtitle=1pt,opacitybacktitle=0.4]

\textbf{Answer to RQ5:} 
In repositories containing previously unidentified defects, \name suggests 61.49\% precise logging fixes according to human evaluation, illustrating its real-world practicality. While this demonstrates considerable effectiveness on unseen data, it also highlights that expert validation remains an essential part of the workflow, although for these successful cases, the effort is largely reduced to reviewing the proposed fix. Furthermore, out of 40 changes reported, developers have merged 25 cases, further validating its usefulness.

\end{tcolorbox}

\subsection{Case Study}
Table~\ref{tab:rq5} shows five representative real-world defective logging statements updated by \name (four successful cases and one failure case). 
In the first four success cases:
(1) The statement-code inconsistency merged by infinispan~\cite{statement_code_2024} fixes the misleading as the older logging statement implies that a commit operation was undertaken, whereas the transaction actually involves rolling back. 
(2) In the static-dynamic inconsistency case merged by trino~\cite{static_dynamic_2024}, the original statement logged variable \textit{sortChannels} twice, mistakenly applying it to record both the channels and orders in the message.
(3) For the temporal inconsistency case merged by openhab-addons~\cite{temporal_inconsistency_2024}, The old statement used the past tense (``\textit{started}''), suggesting that the thread had already started. However, in reality, the thread initiation occurs later in the code execution.
(4) In the readability issue merged by openhab-addons~\cite{readability_issues_2024}, the original one contains a typo \textit{Intellflo}. \name successfully update it to \textit{IntelliFlo} (a pool pump) rather than \textit{Intelliflo} (a software platform).
These real-world updates, which were confirmed by developers, indicate the practicality of \name in detecting and fixing various defective logging statements.

The last row shows a failure (false positive) case offered by \name, where it erroneously reports the logging statement as defective due to a misinterpretation of \textit{next+1} as the current number.  
This error highlights \name's difficulty in handling numerical variable changes, likely due to the limited mathematical capabilities of LLMs. Despite several false positive cases, \name has significantly reduced the manual effort required to inspect all logging statements in a project, enhancing development efficiency and reliability. To eliminate these issues in the future direction, an advanced program logical comprehension module could be introduced.
\section{Discussion}
\label{discussion}
\subsection{Advantages of \name}

The effectiveness of \name has been evidenced by the new project dataset, with 25 out of 40 suggested modifications being accepted and integrated into widely-used, real-world repositories. We expand upon \name in the context of actual logging practices.

\subsubsection{Benefits of utilizing Checker component}
The Checker is designed to mitigate false positives from the similarity-based detector, which are prevalent due to the highly imbalanced nature of real-world data (i.e., most logging statements are not defective). By comparing the effectiveness of Checker in both balanced data and unbalanced data, we argue that Checker is quite beneficial for the actual development scenario. For the balanced dataset, Figure 5 shows a noticeable improvement in precision (increases of 0.15 and 0.11) with a modest reduction in recall, confirming its role in reducing false positives as intended. The value of the Checker emerges in unbalanced datasets. For example, in our new project dataset with 100K entries, the similarity-based classifier initially identified 1,771 potential defective logging statements, and only 149 of them remained after passing Checker. We further manually checked 100 randomly selected cases considered false positives by the Checker and revealed only 4 false negatives, demonstrating a 96\% filtering accuracy while saving the cost of LLM invocation in the subsequent step.

\subsubsection{Applicability of automatically generated logging statements}
\label{applicability_auto}
Although recent works have investigated the capabilities of LLMs in generating logging statements automatically~\cite{li2023exploring,li2024go}, \name continues to offer unique benefits from the following two aspects. First, the generated logging statements is far from mature. LLMs exhibit potential limitations, particularly in accurately generating static text intertwined with dynamic variables — a key area prone to defects~\cite{li2023exploring}. For example, the logging statement generated by the SOTA solution~\cite{li2024go} achieved a ROUGE-L score of 0.509. However, the ROUGE-L score comparing the logging statements before and after commit changes was 0.707. The performance of LLMs in automatically writing logging statements is still far from being applicable, let alone enabling them to avoid various complex defects. Our \name is designed to address these intricacies, identifying and rectifying such defects. Second, while LLM-based generation approaches are effective in generating new logging statements to some extent, they are not typically utilized to retrospectively analyze and update large amounts of existing logging statements within codebases due to computational and practical constraints. \name fills this gap by efficiently processing vast quantities of existing logging statements, ensuring they meet current standards without significant overhead.

\subsubsection{Capability to solve multiple defects in one logging statement situation}
During our manual analysis, we find one defective logging statement case (1/258, 0.3\%) containing multiple defects—specifically, a combination of statement-code inconsistency and readability issues. This finding indicates that it is indeed quite rare for a logging statement to simultaneously exhibit multiple types of defects. Our approach is capable of addressing related issues. The detection model still provides classification results for these special cases, following which Updater performs updates based on the corresponding prompt.

\subsubsection{Capability for extending new defect categories}
In this study, we conducted a manual analysis of log-centric commits sampled from 641 repositories, identifying four distinct defect types, which collectively account for 51.6\%—over half of the cases. It is possible that specific repositories may exhibit unique new defect types. Our proposed method is well-suited to adapt to such scenarios. By retraining the similarity-based classifier and making minor adjustments to the prompt, our approach can seamlessly extend to incorporate one or more additional defect categories.

\subsubsection{Cost-management for \name's pipeline design}
\name's architecture prioritizes cost-effectiveness, particularly in environments with numerous logging statements. By separating the detection and updating phases, \name efficiently filters out non-defective logging statements, minimizing the need for expensive LLM invocations. With a CPI of 0.0011, \name is approximately 16.2 times more cost-efficient than the baseline model GPT4o (CPI of 0.0194). This significant cost reduction, achieved while maintaining superior performance metrics, is primarily due to fewer API calls required for processing. In practical scenarios, the potential savings could exceed those measured in RQ2, as most logging statements do not need LLM processing. By focusing only on relevant context, \name reduces the information sent to the LLM, enhancing cost-efficiency. This makes \name a compelling solution for practitioners seeking to improve logging practices without incurring high costs.

\subsection{Threats to Validity}
\subsubsection{Construct validity}
\textbf{Fine-tuning dataset construction.}
The fine-tuning dataset construction process may raise concerns. Firstly, similar to logging-related works~\cite{mastropaoloUsingDeepLearning2022,liWhereShallWe2020,liuTeLLLogLevel2022}, the basic assumption is that the logging statements in the latest programs are defect-free. Secondly, due to the required fine-tuning set scale, manually labeling a large number of negative examples would be impractical. Consequently, we employed mutation techniques to automatically synthesize defective examples, potentially leading to the inclusion of suboptimal negative examples. 
To mitigate this, the selection of data source is limited to well-maintained codebases (details in \ref{wmr}), which are considered as positive data. Additionally, we devised various mutation methods to ensure the diversity, realism,  and scalability of synthetic data.
\textbf{Limitation in scope of defect types.} While our work focuses on four factually verifiable defect types (e.g., semantic inconsistencies), we intentionally excluded logging-level changes (24.8\%) and other subjective practices. This exclusion introduces a threat to the method's comprehensiveness, as misuse logging-level issues are critical to system observability. However, our decision aligns with the prioritization of objective, context-agnostic defects to build a robust automated framework. To mitigate this threat, we plan to enhance our framework by supporting context-aware misuse log-level detection in the future work.
\textbf{Repair Effectiveness Measurement.}
A potential concern pertains to the practical implication of the 61.49\% success rate reported in RQ5. We acknowledge that this rate, while demonstrating the tool's capability, signifies that manual intervention by domain experts is still necessary to validate the suggestions and correct the remaining 38.51\% of cases where the detection or the proposed fix might be inaccurate. It is important to clarify that LogUpdater is designed to significantly reduce and shift the manual workload associated with logging statement maintenance, rather than eliminating it entirely. Compared to fully manual inspection or approaches relying solely on defect detection tools (which necessitate manual diagnosis and fixing for 100\% of identified potential issues), LogUpdater offers substantial effort savings. Firstly,  for the 61.49\% successful cases, the developer effort is reduced from diagnosis and correction to primarily validation of the automatically generated fix. Secondly, even for the unsuccessful suggestions, the tool focuses developer attention on specific problematic statements and provides a starting point for correction. Therefore, while human oversight remains crucial, LogUpdater represents a significant step towards automating log quality improvement by handling a substantial portion of the detection and repair tasks, thereby reducing the overall manual burden.

\subsubsection{Internal validity} \textbf{Potential data leakage.} We conduct experiments using various LLM backbones. Because the training data for these models are undisclosed, there is a risk of potential data leakage. To mitigate this, we initiate our data collection in April 2024, postdating the training of most LLMs. However, we recognize that for the existing logpatches dataset, there may be some risk of data leakage when testing the updating effects. To mitigate data leakage, we add synthetic test data and new repositories data. The synthetic test data are generated by the proposed synthetic strategies, which can ensure that LLMs cannot include these data as part of their training dataset. For new repositories, code snippets are current as of April 2024. The LLM updates defective logging statements without relying on April 2024 information, reducing data leakage risk. Experimental results demonstrate that \name can achieve a 61.49\% success rate with new repository data, proving our method's effectiveness.
\textbf{Human involvement.}
We conduct manual studies to identify defects in logging statements and human evaluation of new repositories. This human-involved process may introduce bias during annotation. To minimize this threat, we engage two independent annotators with expertise in the field, and ensure that any discrepancies are resolved through discussion until a consensus is reached. Besides, we get 0.78 Cohen's Kappa for manual defect identification and 0.84 for human evaluation, which is a substantial level of agreement.

\subsubsection{External validity}
\textbf{The selection of programming languages.}
We source data from 641 Java projects, which may impact the generalizability of our method for other languages. Following previous study~\cite{ding2023Temporal}, we choose Java because it is a mature language with multiple built-in logging and third-party logging frameworks. Nevertheless, applying \name to other programming languages may affect its performance because of distinct language features.
While fine-tuning detectors on other language-specific datasets is a promising solution, we aim to incorporate more languages in future work.
\textbf{The selection of LLM backbone.} We select GPT4o as our basic LLM backbone and employ six different popular instruction-taken LLMs for experiments to demonstrate the effectiveness of \name. However, different LLM models and model versions exhibit varying levels of comprehension regarding prompts, leading to fluctuations in model performance during experiments. To ensure reproducibility, we set the temperature to 0, theoretically allowing the model to return the same results for identical queries. Additionally, we plan to extend the relevant experiments to newer emerging LLMs to further assess the generalizability of our approach.
\section{Related Work}
\label{related_work}
\subsection{Studies on Maintaining High-quality Logging Statement} 

In numerous related studies, researchers have concentrated on identifying potential problems associated with logging statements. Li et al.~\cite{li2023Are} dedicated their research to analyzing the readability of log messages within these logging statements, formulating three dimensions pertinent to readability. They delved into the feasibility of implementing automatic classification systems for the readability of log messages. Conversely, Ding et al.~\cite{ding2023Temporal} embarked on a pioneering investigation into the temporal relationship between logging statements and their corresponding source code. Their efforts culminated in the development of a tool designed to pinpoint and address temporal inconsistencies. In a different vein, Bouzenia et al.~\cite{bouzenia2023When} directed their focus towards detecting discrepancies between conditions and logging statements, unveiling CMI-Finder—a model grounded in neural networks capable of identifying inconsistencies in condition-statement pairs. Zhang et al.~\cite{zhang2022studying} undertook a comprehensive analysis of 21 open-source projects, encompassing 70,000 logging statements, which include 48,000 production logging statements and 22,000 test logging statements. This study provided developers and researchers with substantive insights into the interplay between test and production logging. Chen et al.~\cite{chen2019extracting,chen2017Characterizinga} explored and pinpointed anti-patterns manifested in logging statements, incorporating six of these anti-patterns into a static code analysis tool to unearth previously undetected anti-patterns in source code. Additionally, Li et al.~\cite{liDLFinderCharacterizingDetecting2019,li2021studying} pursued an in-depth investigation into duplicate logging statements. By meticulously examining over 3,000 duplicate logging statements and their corresponding code across four expansive systems, they delineated five patterns indicative of duplicate logging smells. Furthermore, they introduced DLFinder, a tool engineered to automatically identify problematic duplicate logging code smells.

The primary distinction between our work and the aforementioned studies lies in the fact that while these studies solely detected specific categories, our approach not only identifies these issues but also provides solutions for their rectification.

\subsection{Empirical Studies on Existing Logging Practices}
Owing to advancements in logging within software engineering, there has been a surge in research exploring logging practices.
Li et al.~\cite{liQualitativeStudyBenefits2021} conducted an in-depth qualitative analysis of both the benefits and limitations of logging in software development.  
Zhou et al.~\cite{zhou2020mobilogleak} delved into the ramifications of logging practices on data leaks within mobile apps. 
Kabinna et al.~\cite{kabinna2018examining} identified that aspects like bug corrections, feature augmentations, and code restructuring often led to modifications in logging statements.
Recently, Li et al.~\cite{li2023exploring} pioneered a study on LLMs for logging statement generation and discovered that employing prompt-based strategies with zero-shot or few-shot learning enhances the generalization capabilities of LLMs.
Zhao et al.~\cite{zhao2023studying} conducted an empirical study on the IDs in logging statements, proposed a simple approach to inject IDs in logging statements to mitigate information loss issues, and studied the information gained by injecting IDs into logs.
Zeng et al. \cite{zengStudyingCharacteristicsLogging2019} and Chen \cite{chen2017Characterizinga} extended the study of Yuan et al. \cite{yuanCharacterizingLoggingPractices2012} to Android and Java systems, finding a massive presence of log statements in the analyzed systems. 
Li et al.~\cite{li2023did} investigate the characteristics of dynamic variables and their importance in practice, and then explore the potential of a variable-aware log abstraction technique.
Lai et al.~\cite{lal2015two} perform an analysis of logging code constructs at two levels. They answer nine research questions related to statistical and content analysis at file level and block level.
Additionally, significant studies have been made in determining log levels~\cite{liuTeLLLogLevel2022,liDeepLVSuggestingLog2021,tang2022automated} and content~\cite{dingLoGenTextAutomaticallyGenerating2022,liuWhichVariablesShould2019}.

\subsection{Logging Statement Automation}
Conventionally, the process of logging statement automation can be categorized into two stages~\cite{chen2021survey,he2021survey}: identifying logging locations and creating logging statements. We refer to these steps as where-to-log and what-to-log~\cite{li2024go}. To address the challenge of determining where to log, researchers have explored various methods~\cite{jiaSMARTLOGPlaceError2018,liWhereShallWe2020,yaoLog4PerfSuggestingLogging2018,zhaoLog20FullyAutomated2017,zhuLearningLogHelping2015,yuanBeConservativeEnhancing2012,liStudyingSoftwareLogging2018}. to identify suitable logging locations within the source code. 
In terms of what to log, the generation of logging statements is typically broken down into three subtasks: generating logging text~\cite{dingLoGenTextAutomaticallyGenerating2022,mastropaoloUsingDeepLearning2022,xu2024unilog,xie2024fastlog}, selecting logging variables~\cite{liuWhichVariablesShould2019,yuanImprovingSoftwareDiagnosability2012,dai2022reval}, and predicting the logging level~\cite{heng2024studying,liWhichLogLevel2017,liDeepLVSuggestingLog2021,liuTeLLLogLevel2022,mizouchiPADLADynamicLog2019}.
The most recent approach (Unilog~\cite{xu2024unilog}, Fastlog~\cite{xie2024fastlog}, SCLogger~\cite{li2024go}, LANCE~\cite{mastropaoloUsingDeepLearning2022}) provides automatically logging statements generation solution of deciding logging places, statements level, content, and variables at one-step by leveraging the ability of LLMs. 

As we detailed discussion in Section~\ref{applicability_auto}, although recent works have investigated the capabilities of LLMs in generating logging statements automatically, \name continues to offer unique benefits by lower costs and better performance in detecting and updating defective logging statements.
\section{Conclusion and Future Work}
\label{conclusion}

In this paper, we find that existing research on analyzing existing logging statements is limited, primarily focusing on detecting a singular type of defect (limited scope of analysis) and relying on manual intervention for fixes rather than automated solutions (detection without repairing).

First, we conduct a broad-scope pilot study and identify four types of common defects that developers are concerned about from 641 targeted repositories. This coverage enables us to analyze defective logging practices across varied development environments, unveiling new insights into developer priorities and defect patterns that are more broadly representative of real-world software development.

Then, we propose a two-phase end-to-end framework to automatically detect and update potential defective logging statements. This is a pioneering advancement in the field, especially valuable for large-scale projects where manual repair is not feasible. By automating repairs, our tool saves significant manual effort, addressing a gap not covered by existing literature. Extensive experiment results demonstrate the effective detection and updating capabilities and practicality of \name. 

In future work, we will extend LogUpdater to handle log-level optimization using context-aware techniques, addressing project-specific conventions while maintaining the framework’s reliability. Specifically, a potential path is to develop a context-aware detection framework combining organizational guidelines and runtime execution frequency. By learning cross-project commit histories and logging statements, the model will learn optimal level assignments that balance observability needs with overhead constraints. The framework will adaptively detect misuse log-level based on code context, enabling project-specific yet automated existing logging statement checking.

\section*{Acknowledgement}
The work described in this paper was supported by the Research Grants Council of the Hong Kong Special Administrative Region, China (No. CUHK 14206921 of the General Research Fund).
\bibliographystyle{ACM-Reference-Format}
\bibliography{bibfile}


\end{document}